\documentclass[12pt,letter]{article}

\usepackage{graphicx, epsfig, color}
\def\eslt{E_T^{\rm miss}}
\def\to{\rightarrow}

\def\bi{\begin{itemize}}
\def\ei{\end{itemize}}
\def\te{\tilde e}

\def\tu{\tilde u}
\def\tc{\tilde c}

\def\tb{\tilde b}

\def\tst{\tilde t}

\def\ttau{\tilde \tau}
\def\tmu{\tilde \mu}
\def\tg{\tilde g}
\def\tnu{\tilde\nu}
\def\tell{\tilde\ell}
\def\tq{\tilde q}
\def\tw{\widetilde W}
\def\tz{\widetilde Z}
\def\alt{\stackrel{<}{\sim}}
\def\agt{\stackrel{>}{\sim}}
\def\be{\begin{equation}}  
\def\ee{\end{equation}}  
\def\bea{\begin{eqnarray}}  
\def\eea{\end{eqnarray}}  
\def\beas{\begin{eqnarray*}}  
\def\eeas{\end{eqnarray*}}  
\newcommand\prd[3]{{\it Phys.\ Rev.\ }{\bf D #1} (#2) #3}

\newcommand\prl[3]{{\it Phys.\ Rev.\ Lett.\ }{\bf #1} (#2) #3}
\newcommand\plb[3]{{\it Phys.\ Lett.\ }{\bf B #1} (#2) #3}
\newcommand\jhep[3]{{\it J. High Energy Phys.\ }{\bf #1} (#2) #3}

\newcommand\npb[3]{{\it Nucl.\ Phys.\ }{\bf B #1} (#2) #3}
\newcommand\epjc[3]{{\it Eur.\ Phys.\ J. }{\bf C #1} (#2) #3}

\newcommand\zpc[3]{{\it Z.\ Phys.\ {\bf C}}{\bf  #1} (#2) #3}
\newcommand\mpla[3]{{\it Mod.\ Phys.\ Lett.\ }{\bf A #1} (#2) #3}
\newcommand{\hepph}[1]{hep-ph/#1}

\newcommand{\hepex}[1]{hep-ex/#1}

\begin{document}
\begin{titlepage}

\vspace{0.5cm}
\begin{center}
{\Large \bf 
Prospects for Hypercharged Anomaly Mediated \\SUSY Breaking
at the LHC
}\\ 
\vspace{1.2cm} \renewcommand{\thefootnote}{\fnsymbol{footnote}}
{\large Howard Baer$^{1}$\footnote[1]{Email: baer@nhn.ou.edu },
Radovan Derm\' \i\v sek$^2$\footnote[2]{Email: dermisek@indiana.edu}, 
Shibi Rajagopalan$^1$\footnote[3]{Email: shibi@nhn.ou.edu},
Heaya Summy$^1$\footnote[4]{Email: heaya@nhn.ou.edu}} \\
\vspace{1.2cm} \renewcommand{\thefootnote}{\arabic{footnote}}
{\it 
1. Dept. of Physics and Astronomy,
University of Oklahoma, Norman, OK 73019, USA \\
2. Dept. of Physics,
Indiana University, Bloomington IN 47405, USA \\
}

\end{center}

\vspace{0.5cm}
\begin{abstract}
\noindent 
We investigate the phenomenological consequences of string models
wherein the MSSM resides on a D-brane, and the hypercharge gaugino mass
is generated in a geometrically separated hidden sector.
This hypercharged anomaly-mediated SUSY breaking (HCAMSB) model 
naturally solves the tachyonic slepton mass problem 
endemic to pure AMSB scenarios.
In HCAMSB, one obtains a mass ordering $M_1>\mu >M_2$ with split left- and right- scalars, 
whereas in mAMSB models, one obtains $\mu >M_1>M_2$ with nearly degenerate left- and right-
scalars. 
We compute the allowed parameter space and 
expected superparticle mass spectrum in the HCAMSB model. 
For low values of the HC and AMSB mixing parameter $\alpha$,
the spectra is characterized by light left-sleptons, while the spectra
for large $\alpha$ is characterized by light top- and bottom-
squarks. 
We map out the approximate reach of LHC for HCAMSB, and
find that with 100 fb$^{-1}$ of integrated luminosity, a gravitino mass of
$\sim 115$ (105) TeV can be probed for low (high) values of $\alpha$, 
corresponding to a gluino mass reach of $\sim 2.4$ (2.2) TeV.
Both cases contain-- as is typical in AMSB models-- long lived charginos
that should yield visible highly ionizing tracks in the LHC detector.
Also, in the lower $\tan\beta$ range, HCAMSB models give rise to reconstructable
$Z\to \ell\bar{\ell}$ candidates in SUSY cascade decay events, 
while mAMSB models should do so only rarely.
\vspace*{0.8cm}


\end{abstract}


\end{titlepage}

\section{Introduction}
\label{sec:intro}

Anomaly-mediated supersymmetry breaking (AMSB) models have received much
attention in the literature due to their attractive properties\cite{amsb}:
the soft supersymmetry (SUSY) breaking terms are completely 
calculable in terms of just one free parameter (the gravitino mass, $m_{3/2}$),
the soft terms are real and flavor invariant, thus solving the SUSY flavor and $CP$ 
problems, the soft terms are actually renormalization group invariant\cite{jj}, 
and can be 
calculated at any convenient scale choice. In order to realize the AMSB set-up,
the hidden sector must be ``sequestered'' on a separate brane 
from the observable sector in an extra-dimensional
universe, so that tree-level supergravity breaking terms do not dominate the 
soft term contributions. 
Such a set-up can be realized in brane-worlds, where SUSY breaking takes
place on one brane, with the visible sector residing on a separate brane. 
The soft SUSY breaking (SSB) terms arise from the rescaling anomaly.

In spite of its attractive features, AMSB models suffer from the well-known problem
that slepton mass-squared parameters are found to be negative, giving rise to tachyonic states.
The original solution to this problem is to suppose that scalars acquire as well a 
universal mass $m_0$, which when added to the AMSB SSB terms, renders them positive.
Thus, the parameter space of the ``minimal'' AMSB model (mAMSB) is given by
\be
m_0,\ m_{3/2},\ \tan\beta ,\ sign(\mu ) .
\ee

An alternative set-up for AMSB has been advocated in Ref. \cite{dvw},
known as hypercharged anomaly-mediation (HCAMSB).
It is a string motivated scenario which uses a similar setup as the one 
envisioned for AMSB.
In HCAMSB, SUSY breaking is localized at the bottom of a strongly warped hidden region,
geometrically separated from the visible region where the MSSM resides.
The warping suppresses contributions due to tree-level gravity mediation\cite{Kachru:2007xp} 
and the anomaly mediation\cite{amsb} can become the dominant 
source of SUSY breaking in the visible sector. 
Possible exceptions to this sequestering mechanism are 
gaugino masses of $U(1)$ gauge symmetries~\cite{rrform}. 
Thus, in the MSSM, the mass of the bino-- the gaugino  of $U(1)_Y$-- can be the 
only soft SUSY breaking parameter not determined by anomaly mediation\cite{dvw}. 
Depending on its size, 
the bino mass $M_1$ can lead to a small perturbation to the spectrum of anomaly mediation, 
or it can be the largest soft SUSY breaking parameter in the visible sector: 
as a result of RG evolution its effect on other soft SUSY breaking parameters can 
dominate the contribution from anomaly mediation.
In extensions of the MSSM, additional $U(1)'$s can also communicate SUSY breaking 
to the MSSM sector~\cite{Langacker:2007ac}.

Besides sharing the same theoretical setup, anomaly mediation and hypercharge mediation cure 
phenomenological shortcomings of each other. 
The minimal AMSB model predicts a negative mass squared for the sleptons 
(and features relatively heavy squarks). On the other hand, the pure hypercharge mediation 
suffers from negative squared masses for stops and sbottoms 
(and features relatively heavy sleptons): see Sec. \ref{sec:pspace}.
As a result, the combination of hypercharge and anomaly mediation leads to 
phenomenologically viable spectra
for a sizable range of relative contributions~\cite{dvw}.

We parametrize the HCAMSB SSB contribution $\tilde{M}_1$ using a dimensionless quantity 
$\alpha$ such that $\tilde{M}_1 =\alpha m_{3/2}$ so that $\alpha$ 
governs the size of the hypercharge
contribution to soft terms relative to the AMSB contribution. 
Then the parameter space of HCAMSB models is given by
\be
\alpha ,\ m_{3/2},\ \tan\beta ,\ sign(\mu ) .
\ee
In the HCAMSB model, we assume as usual that electroweak symmetry is broken radiatively
by the large top-quark Yukawa coupling. Then  the SSB $B$ term and the superpotential $\mu$
term are given as usual by the scalar potential minimization conditions which 
emerge from requiring an appropriate breakdown of electroweak symmetry. 

In HCAMSB, we take the SSB terms to be of the form:
\bea
M_1 &=& \tilde{M}_1+\frac{b_1g_1^2}{16\pi^2}m_{3/2},\cr
M_a &=& \frac{b_a g_a^2}{16\pi^2}m_{3/2},\ \ \ a=2,\ 3\ \ \cr
m_i^2 &=& -{1\over 4}\left\{ \frac{d\gamma}{dg}\beta_g+\frac{d\gamma}{df}\beta_f\right\}m_{3/2}^2 \cr
A_f &=& \frac{\beta_f}{f}m_{3/2} ,
\eea
where $(b_1,b_2,b_3)=(33/5,1,-3)$, $\beta_f$ is the beta function for the corresponding
superpotential coupling, and $\gamma =\partial\ln Z /\partial\ln\mu$ with $Z$ the wave function
renormalization constant.
The wino and gluino masses ($M_2$ and $M_3$) receive a contribution from the bino mass at 
the two loop level. 
Thus, in pure hypercharge mediation, they are one loop suppressed compared to the scalar masses.
For convenience, we assume the above SSB mass parameters are input at the GUT scale, 
and all weak scale SSB parameters are determined by renormalization group evolution.

We have included the above HCAMSB model into the Isasugra subprogram of the 
event generator Isajet v7.79\cite{isajet}. After input of the above parameter set, Isasugra
then implements an iterative procedure of solving the MSSM RGEs for the
26 coupled renormalization group equations, taking the weak scale 
measured gauge couplings and third generation Yukawa couplings as inputs, as well
as the above-listed GUT scale SSB terms. Isasugra implements full 2-loop RG running
in the $\overline{DR}$ scheme, and minimizes the RG-improved 1-loop effective
potential at an optimized scale choice $Q=\sqrt{m_{\tst_L}m_{\tst_R}}$\cite{hh} 
to determine the magnitude of $\mu$ and $m_A$. All physical sparticle masses
are computed with complete 1-loop corrections, and 1-loop weak scale threshold corrections
are implemented for the $t$, $b$ and $\tau$ Yukawa couplings\cite{pbmz}. The off-set of the 
weak scale boundary conditions due to threshold corrections (which depend on the entire
superparticle mass spectrum), necessitates an iterative up-down RG running solution.
The resulting superparticle mass spectrum is typically in close accord with other
sparticle spectrum generators\cite{kraml}.

Once the weak scale sparticle mass spectrum is known, then sparticle 
production cross sections and branching fractions may be computed, and
collider events may be generated. Then, signatures for HCAMSB at the CERN LHC may be
computed and compared against Standard Model (SM) backgrounds.
Our goal in this paper is to characterize the HCAMSB parameter space and
sparticle mass spectrum, and derive consequences for the CERN LHC $pp$ collider,
which is expected to begin operation in Fall, 2009. Some previous 
investigations of mAMSB at LHC have been reported in Ref. \cite{amsb_lhc,bmt,barr}.

The remainder of this paper is organized as follows. In Sec. \ref{sec:pspace},
we calculate the allowed parameter space of HCAMSB models, imposing various
experimental and theoretical constraints. We also show sample mass spectra from HCAMSB models,
and show their variation with $\alpha$ and $m_{3/2}$. We show typical values of 
$BF(b\to s\gamma )$ and $(g-2)_\mu$ that result.
In Sec. \ref{sec:lhc}, we explore consequences of the HCAMSB model for LHC sparticle
searches. Typically, collider events are characterized by production of
high $p_T$ $b$ and $t$ quarks, along with $\eslt$ and observable tracks from
late decaying charginos $\tw_1^\pm$. 
For small $\alpha$, slepton pair production may be visible, 
while for large $\alpha$, direct $\tst_1\bar{\tst}_1$ and $\tb_1\bar{\tb}_1$ production 
may be visible. The LHC reach for 100 fb$^{-1}$ should extend up to
$m_{3/2}\sim 115$ (105) TeV, corresponding to a reach in $m_{\tg}\sim 2.4$ (2.2) TeV,
for small (large) values of $\alpha$.
The HCAMSB model should be easily distinguishable from the mAMSB model 
at the LHC if $\tan\beta$ is not too large, 
due to the presence of $Z\to\ell\bar{\ell}$ candidates in cascade decay events.
The presence of these reflects the mass ordering $M_1>\mu >M_2$ in the HCAMSB model,
while $\mu  >M_1>M_2$ in the mAMSB model.
In Sec. \ref{sec:conclude}, we present our conclusions and outlook for HCAMSB models.

\section{Mass spectra, parameter space and constraints on the HCAMSB model}
\label{sec:pspace}

\subsection{Spectra and parameter space}

\begin{figure}[htbp]
\begin{center}
\includegraphics[angle=-90,width=0.75\textwidth]{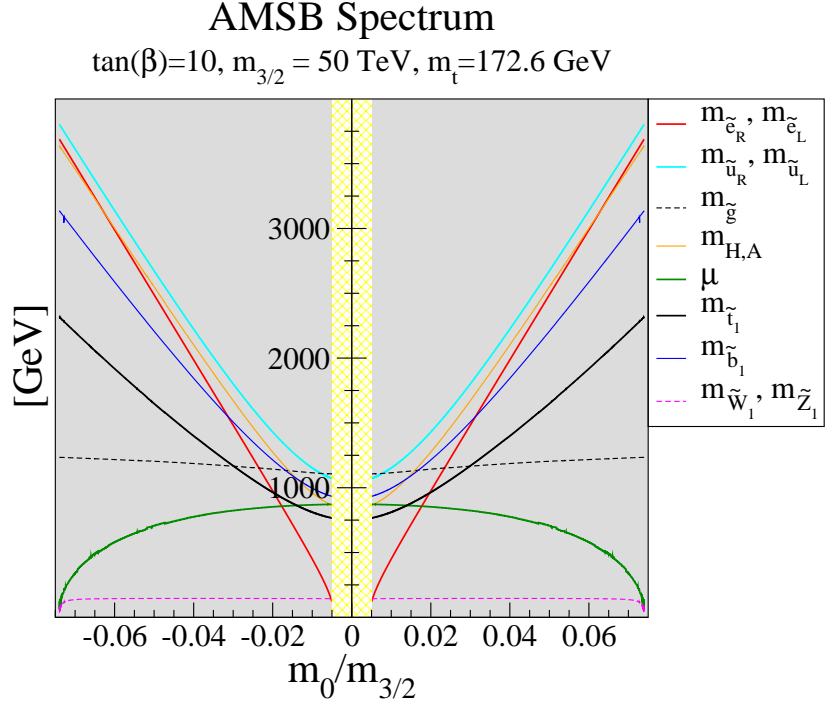}
\includegraphics[angle=-90,width=0.75\textwidth]{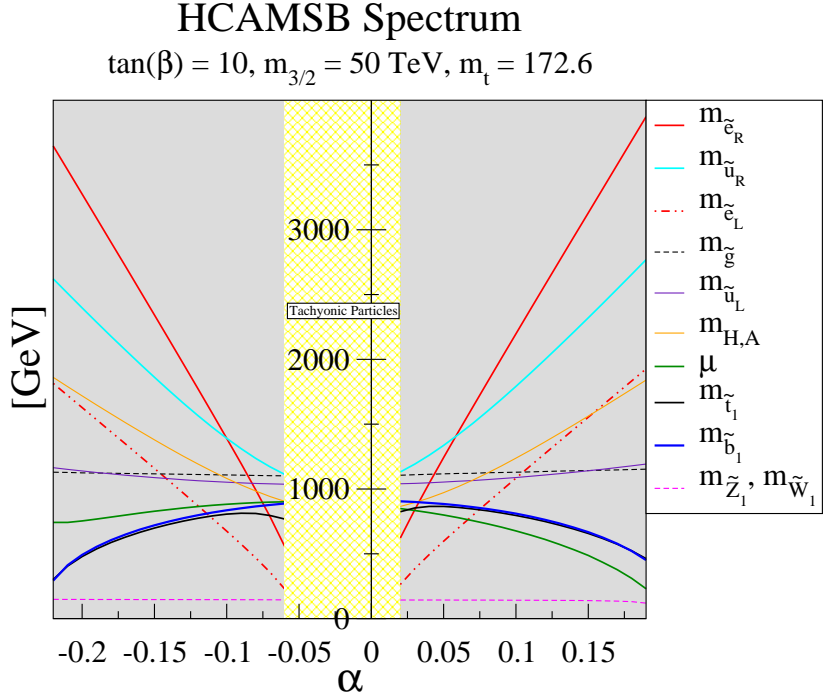}
\vspace{-.5cm}
\caption{Sparticle mass spectrum versus {\it a}). $m_0/m_{3/2}$ in 
mAMSB and {\it b}). $\alpha$ in the HCAMSB model,
for $m_{3/2}=50$ TeV and $\tan\beta =10$, with $\mu >0$ and
$m_t=172.6$ GeV. 
}
\label{fig:m10}
\end{center}
\end{figure}

We begin our discussion by plotting out in Fig. \ref{fig:m10} the mass spectra of
various sparticles versus {\it a}). $m_0/m_{3/2}$ in mAMSB and 
{\it b}). $\alpha$ in the HCAMSB model,  for $m_{3/2}$ fixed at 50 TeV, 
while taking $\tan\beta =10$, $\mu >0$ and $m_t=172.6$ GeV.
For $m_0$ and $\alpha \sim 0$, the yellow-shaded region yields the 
well-known tachyonic slepton mass-squared values, which could lead
to electric charge non-conservation in the scalar potential.
In mAMSB, as $m_0$ increases, all the scalars increase in mass, while
$m_{\tg}$, $m_{\tw_1}$ and $m_{\tz_1}$ remain roughly constant, and the 
superpotential $\mu$ term decreases. The large $m_0$ limit of parameter space is 
reached around $m_0/m_{3/2}\sim 0.075$, where EWSB is no 
longer properly broken (signaled by $\mu^2<0$). 
We also see the well-known property of mAMSB models that 
$m_{\te_L}\simeq m_{\te_R}$.
In addition, an important distinction between the two models is the mass ordering
which enters into the neutralino mass matrix: we find typically that 
$M_1>\mu >M_2$ in the HCAMSB model, while $\mu >M_1>M_2$ in mAMSB.
Thus, both models will have a wino-like $\tz_1$ state. However, in the HCAMSB model, 
the $\tz_{2,3}$ are dominantly higgsino-like states, with $\tz_4$ being bino-like,
while in the mAMSB model, we expect $\tz_2$ to be bino-like with 
$\tz_{3,4}$ being higgsino-like. This mass ordering difference will give rise to
a crucial distinction in LHC SUSY cascade decay events (see Sec. \ref{sec:lhc}) 
which may serve to distinguish the two models.

In the HCAMSB case, as $\alpha$ increases, the GUT scale gaugino mass $M_1$ increases.
Thus, the bino mass increases with $\alpha$, while the light charginos
$\tw_1^\pm$ and neutralino $\tz_1$ remain wino-like with mass fixed near $M_2$,
and the gluino remains with mass fixed at nearly $M_3\sim 0.022 m_{3/2}$.
Many of the scalar masses also vary with $\alpha$. The reason is that as
$\alpha$ increases, so does the GUT scale value of $M_1$. The 
large value of $M_1$ feeds into the scalar masses via their
renormalization group equations, causing many of them to increase
with $\alpha$, with the largest increases
occurring for the scalars with the largest weak hypercharge assignments $Y$.
Thus, we see strong increases in the $\tu_R$, $\te_L$ and
especially the $\te_R$ masses with increasing $|\alpha |$. The $\tu_L$ 
squark only receives a small increase in mass, since its hypercharge value
is quite small: $Y=1/3$. 
From Fig. \ref{fig:m10}{\it b})., we already see an important distinction
between mAMSB and HCAMSB models: in the former case, the
$\te_L$ and $\te_R$ states are nearly mass degenerate, while in the 
latter case these states are highly split, with $m_{\te_R}\gg m_{\te_L}$.

An exception to the mass increase with $\alpha$ in Fig. \ref{fig:m10}{\it b}). 
occurs in the values of $m_{\tst_1}$ and $m_{\tb_1}$. 
In these cases, the large increase
in $m_{U_3}^2$ feeds into the RGE $X_t=m_{Q_3}^2+m_{U_3}^2+m_{H_u}^2+A_t^2$ term\cite{wss}, 
and {\it amplifies} the top-quark Yukawa coupling suppression of the 
$m_{Q_3}^2$ term. 
Since the doublet $Q_3$ contains both the $\tst_L$ and $\tb_L$ states, both of these
actually suffer a {\it decrease} in mass with increasing $\alpha$.
Thus, we expect in HCAMSB models with moderate to large $\alpha$ 
that the third generation squark states will be highly split. For large
$|\alpha |$, we expect the light third generation squarks 
$\tst_1$ and $\tb_1$ to be quite light, with a dominantly left-
squark composition. The heavier squarks $\tst_2$ and $\tb_2$ will
be quite heavy, and dominantly right-squark states. 

\begin{figure}[htbp]
\begin{center}
\includegraphics[width=0.5\textwidth]{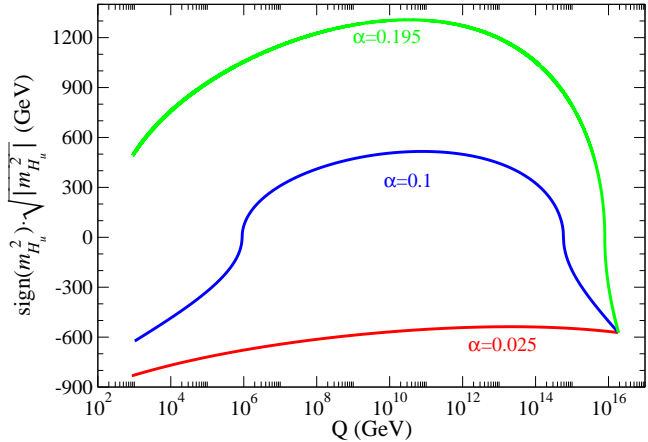}
\caption{Running of the $m_{H_u}^2$ parameter as a function of
energy scale $Q$ for $\alpha =0.025$, $0.1$ and $0.195$ for
$m_{3/2}=50$ TeV and $\tan\beta =10$, $\mu >0$ in the 
HCAMSB model.
}
\label{fig:mhu}
\end{center}
\end{figure}

In addition, we see from Fig. \ref{fig:m10}{\it b}). that the superpotential $\mu$
term {\it decreases} with increasing $\alpha$. At moderate-to-large $\tan\beta$, 
the $\mu$ term is-- from the tree-level scalar potential minimization conditions--
$\mu^2\simeq -m_{H_u}^2$. The running of $m_{H_u}^2$ versus energy scale $Q$ is shown
in Fig. \ref{fig:mhu} for $\alpha =0.025,\ 0.1$ and 0.195. We see that as $\alpha$
increases, the value of $-m_{H_u}^2$ actually decreases, leading to a small $\mu^2$ value.
The relevant RGE reads
\be
\frac{dm_{H_u}^2}{dt} =\frac{2}{16\pi^2}\left(- \frac{3}{5}g_1^2M_1^2 -3g_2^2M_2^2+
\frac{3}{10}g_1^2 S+3f_t^2 X_t\right) .
\ee
A large value of $M_1$ thus leads to an {\it upwards} push to $m_{H_u}^2$
in its early running from $Q=M_{GUT}$, which is only later compensated by the downward
push of the Yukawa-coupling term involving the top Yukawa coupling $f_t$.
In the figure, for the case of $\alpha =0.195$, the weak scale value of
$m_{H_u}^2$ is actually positive. Upon adding the large 1-loop corrections to the
effective potential (due to the light top-squark), the RG-improved scalar potential
yields a positive value of $\mu^2$. Thus, in the region of large
$\alpha$, where $\mu$ becomes small and comparable to $M_2$, we expect the
neutralino $\tz_1$ to become a mixed wino-higgsino particle, and the
corresponding $\Delta m= m_{\tw_1}-m_{\tz_1}$ mass gap to increase beyond the value
$\Delta m\sim 150$ MeV which is expected in AMSB models\cite{matchev}. 

An interesting coincidence related to the RG evolution of $m_{H_u}^2$ in the limit 
where hypercharge mediation dominates is that the electroweak symmetry breaking {\it requires} 
the electroweak scale to be $\sim (12-16)$ orders of magnitude below the  scale $M_{\star}$  
($M_{\star}$ may be of order the GUT scale or string scale) 
at which the Bino mass $M_1$ is generated. 
If the hierarchy between the electroweak scale and $M_{\star}$ was smaller, then a 
SUSY breaking scenario in which hypercharge mediation  dominates would  not be capable of 
triggering EWSB (the energy interval for RG evolution would not be large enough to drive 
the $m_{H_u}^2$ parameter  to negative values). 
This is a very uncommon feature among SUSY breaking scenarios.\footnote{A similar feature
can be found in scenarios with negative stop masses squared 
at the unification scale~\cite{Dermisek:2006ey}. For more details, see 
also Ref. \cite{Dermisek:2009si}.}

%
%

\begin{figure}[htbp]
\begin{center}
\includegraphics[angle=-90,width=.8\textwidth]{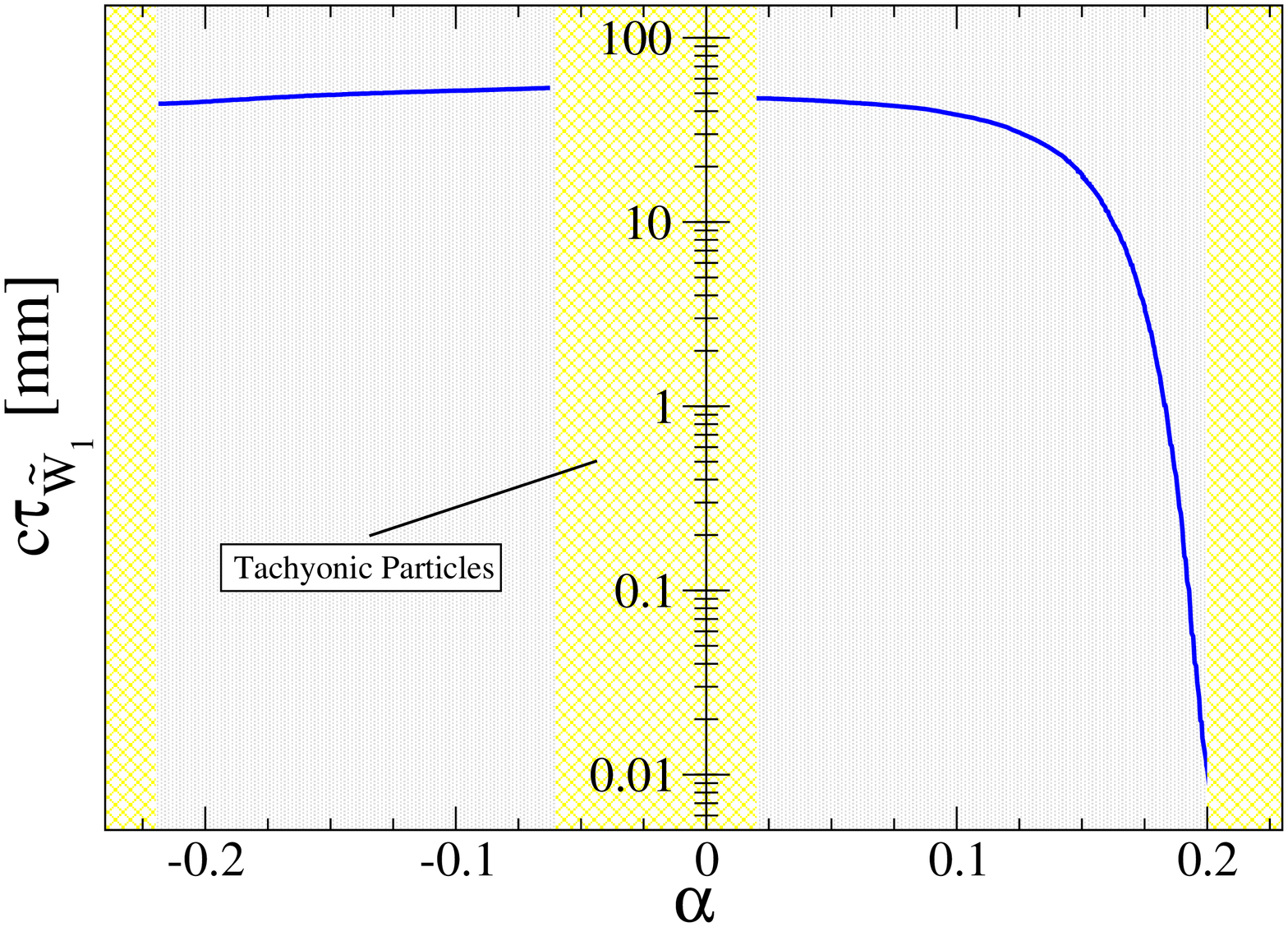}
\caption{Lifetime $c\tau_{\tw_1}$ in $mm$ of the light chargino state
versus $\alpha$ in the HCAMSB model
for $m_{3/2}=50$ TeV, $\tan\beta =10$ and  $\mu >0$.}
\label{fig:ctau}
\end{center}
\end{figure}

For a more detailed comparison, we list in Table \ref{tab:cases}
the sparticle mass spectrum for a mAMSB point with
$m_0=300$ GeV, $m_{3/2}=50$ TeV, $\tan\beta =10$ and $\mu >0$,
and two HCAMSB points with small and large $\alpha$ values
equal to $0.025$ and $0.195$. While all three cases have a comparable
gluino mass, we see that the rather small splitting amongst
$\tu_L -\tu_R$ and also $\te_L -\te_R$ states in mAMSB is turned 
to large left-right splitting in the HCAMSB cases. We also see that the
$m_{\tw_1}-m_{\tz_1}\sim 150$ MeV mass gap in AMSB and HCAMSB1-- which leads to
long-lived and possibly observable $\tw_1$ tracks in collider detectors-- 
opens up to a few GeV in the HCAMSB2 case. 
The latter mass gap is large enough to make the $\tw_1$ state
less long lived, although still maintaining possibly measureable tracks in collider
scattering events. The value of $c\tau_{\tw_1}$ versus $\alpha$ is shown in Fig. \ref{fig:ctau},
where we usually get $c\tau_{\tw_1}\sim 10-100$ mm for most $\alpha$ values. The value drops 
to shorter lengths for large $\alpha$. The shorter travel time of the $\tw_1$ would
distinguish the large $\alpha$ HCAMSB case with a mixed higgsino-wino $\tz_1$ state 
from the low $\alpha$ HCAMSB case where $\tz_1$ is instead nearly pure wino-like.
%
\begin{table}
\begin{center}
\begin{tabular}{lccc}
\hline
parameter & mAMSB & HCAMSB1 & HCAMSB2 \\
\hline
$\alpha$    & --- & 0.025 & 0.195  \\
$m_0$       & 300 & --- & --- \\
$m_{3/2}$   & $50\ {\rm TeV}$ & $50\ {\rm TeV}$ & $50\ {\rm TeV}$ \\
$\tan\beta$ & 10 & 10 & 10 \\
$M_1$       & 460.3   & 997.7 & 4710.5 \\
$M_2$       & 140.0   & 139.5 & 137.5 \\
$\mu$       & 872.8 & 841.8 & 178.8 \\
$m_{\tg}$   & 1109.2 & 1107.6 & 1154.2 \\
$m_{\tu_L}$ & 1078.2 & 1041.3 & 1199.1 \\
$m_{\tu_R}$ & 1086.2   & 1160.3 & 2826.3 \\
$m_{\tst_1}$& 774.9 & 840.9 & 427.7 \\
$m_{\tst_2}$& 985.3 & 983.3 & 2332.5 \\
$m_{\tb_1}$ & 944.4 & 902.6 & 409.0  \\
$m_{\tb_2}$ & 1076.7 & 1065.7 & 1650.7 \\
$m_{\te_L}$ & 226.9 & 326.3  & 1973.1 \\
$m_{\te_R}$ & 204.6 & 732.3  & 3964.9 \\
$m_{\tw_2}$ & 879.2 & 849.4 & 233.1 \\
$m_{\tw_1}$ & 143.9 & 143.5 & 107.1 \\
$m_{\tz_4}$ & 878.7 & 993.7 & 4727.2 \\ 
$m_{\tz_3}$ & 875.3 & 845.5 & 228.7 \\ 
$m_{\tz_2}$ & 451.1 & 839.2 & 188.6 \\ 
$m_{\tz_1}$ & 143.7 & 143.3 & 105.0 \\ 
$m_A$       & 878.1 & 879.6 & 1875.1 \\
$m_h$       & 113.8 & 113.4 & 112.1 \\ 
$\Omega_{\tz_1}h^2$ & 0.0016 & 0.0015 & 0.0011 \\
\hline
$BF(\tz_2\to\tz_1 Z)$ & $0.01\%$ & 7.7\% & 22.3\% \\
\hline
$\sigma\ [{\rm fb}]$ & $7.7\times 10^3$ & $7.4\times 10^3$ & $1.8\times 10^4$ \\
$\tg ,\tq\ pairs$ & 15.0\% & 15.5\% & 14.3\% \\
$EW-ino\ pairs$ & 79.7\% & 81.9\% & 85\% \\
$slep.\ pairs$ & 3.7\% & 0.8\% & -- \\
$\tst_1\bar{\tst}_1$ & 0.4\% & 0.2\% & 5.5\% \\
\hline
\end{tabular}
\caption{Masses and parameters in~GeV units
for three case study points AMSB, HCAMSB1 and HCAMSB2
using Isajet 7.79 with $m_t=172.6$ GeV and $\mu >0$. 
We also list the 
total tree level sparticle production cross section 
in fb at the LHC.
}
\label{tab:cases}
\end{center}
\end{table}

\begin{figure}[htbp]
\begin{center}
\includegraphics[angle=-90,width=.8\textwidth]{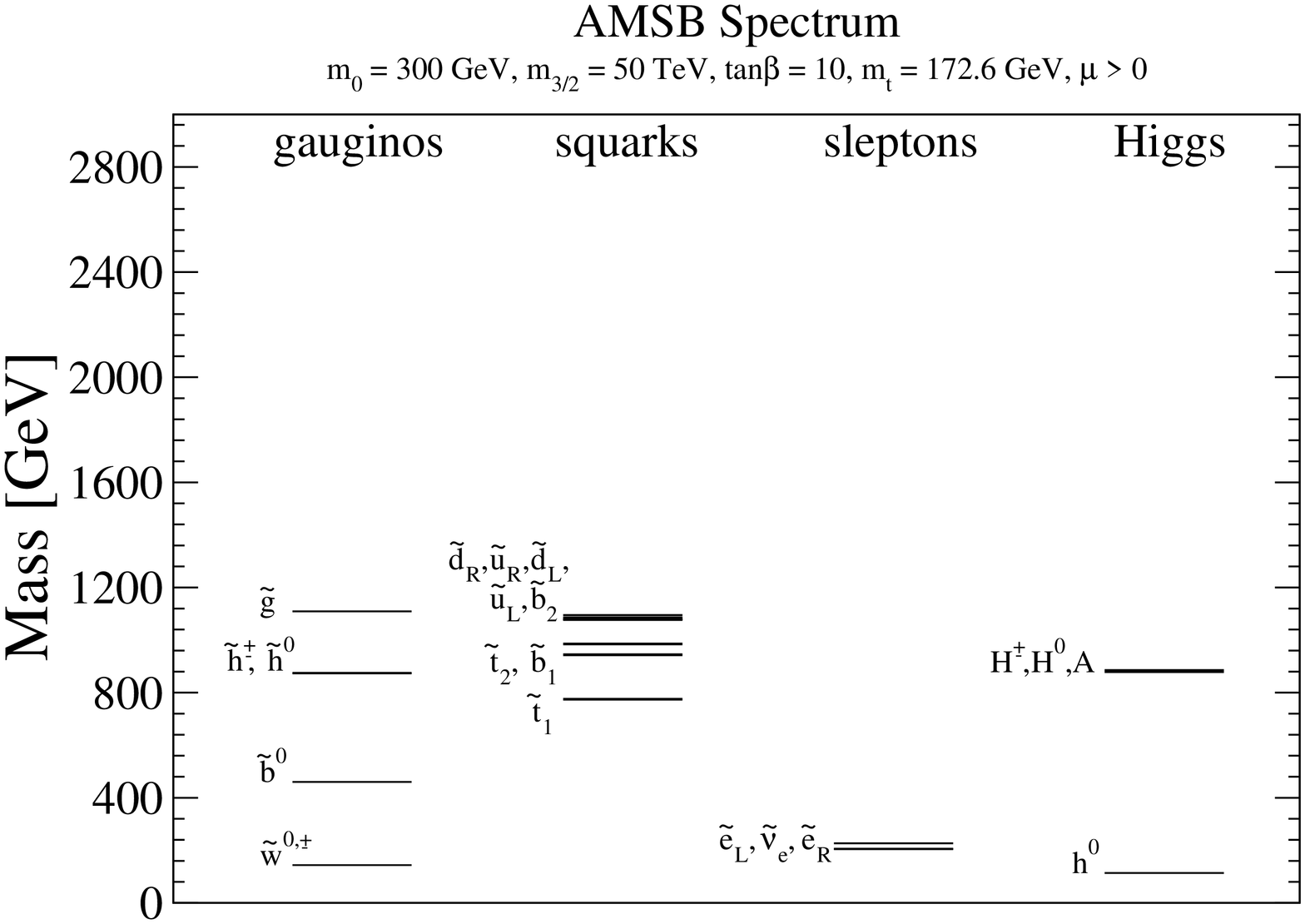}
\includegraphics[angle=-90,width=.8\textwidth]{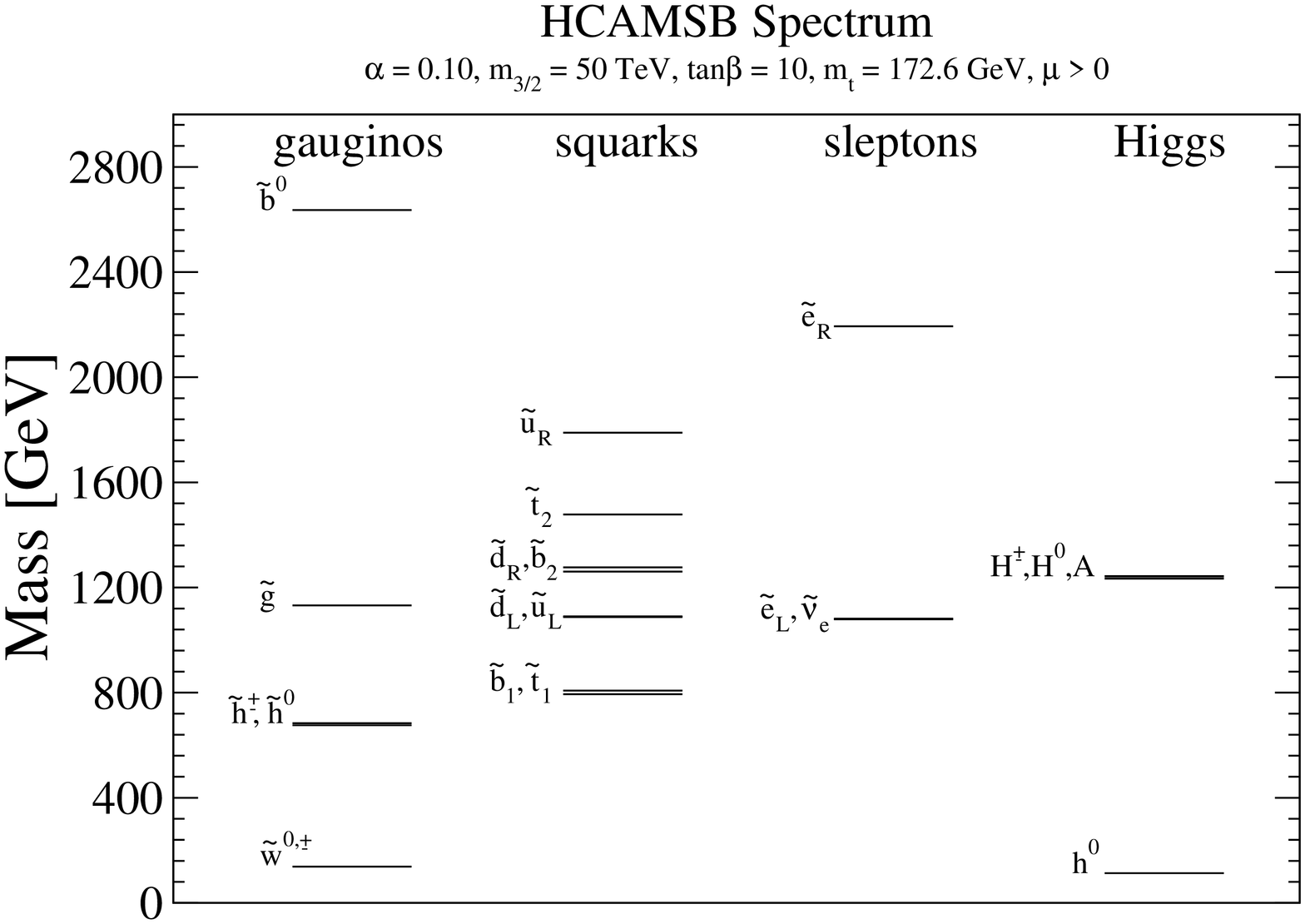}
\caption{Mass spectra for mAMSB and HCAMSB models with
$m_{3/2}=50$ TeV, $\tan\beta =10$ and  $\mu >0$. For mAMSB, we take
$m_0=300$ GeV, while for HCAMSB, we take $\alpha= 0.1$.}
\label{fig:spec}
\end{center}
\end{figure}

We show in Fig. \ref{fig:spec} a cartoon of the mass spectra for mAMSB and 
HCAMSB taken at the same values of $m_{3/2}=50$ TeV, $\tan\beta =10$ and
$\mu >0$. For mAMSB, we take $m_0=300$ GeV, while for HCAMSB, we take 
$\alpha =0.1$. The figure illustrates quickly the main features of a
left-right scalar degeneracy in mAMSB, but a 
left-right split spectrum of HCAMSB models. 
It also illustrates the $\mu>M_1>M_2$ ordering in mAMSB, and
$M_1>\mu >M_2$ in HCAMSB via the location of the wino, 
higgsino and bino states.

\begin{figure}[htbp]
\begin{center}
\includegraphics[angle=-90,width=1.0\textwidth]{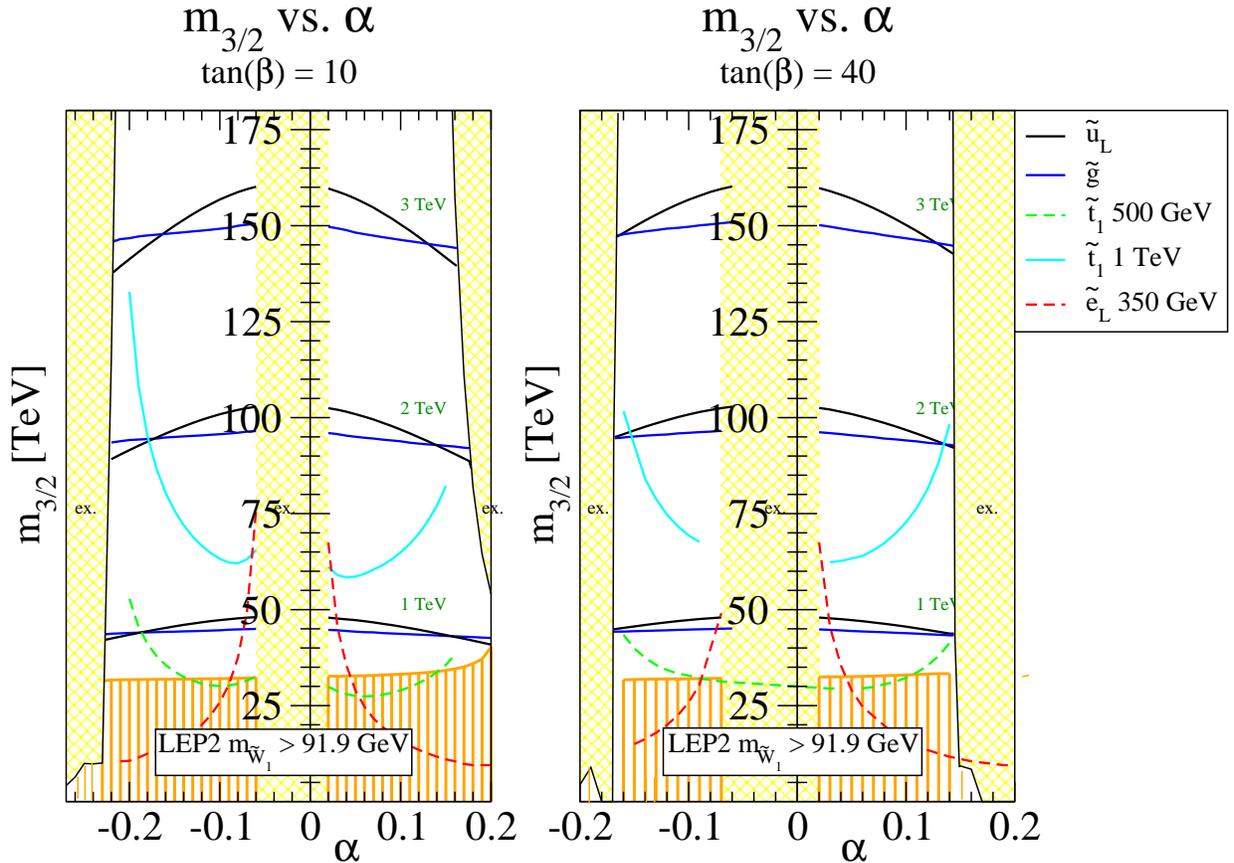}
\caption{Allowed parameter space in the $m_{3/2} \ vs. \ \alpha$ 
plane in the HCAMSB model with $\mu >0$, $m_t=172.6$ GeV and 
{\it a}). $\tan\beta =10$ and {\it b}). $\tan\beta =40$. 
We also show contours of $m_{\tu_L}$, $m_{\tg}$, $m_{\te_L}$ and $m_{\tst_1}$.
}
\label{fig:pspace}
\end{center}
\end{figure}

Next, we display the allowed parameter space of the 
HCAMSB model in the $m_{3/2} \ \ vs. \ \ \alpha$ plane in Fig. \ref{fig:pspace}
for {\it a}). $\tan\beta =10$ and {\it b}). $\tan\beta =40$, where
we also take $\mu >0$ and $m_t=172.6$ GeV. The yellow shaded region
around $\alpha\sim 0$ is dis-allowed because this region generates
tachyonic slepton masses. The large $|\alpha |$ solutions are 
forbidden due to a lack of appropriate breakdown of electroweak 
symmetry (here signaled by a superpotential term $\mu^2<0$).
Over most of parameter space, the lightest SUSY particle is the
wino-like neutralino $\tz_1$, although for large $|\alpha |$, 
the $\tz_1$ becomes a mixed higgsino-wino state (due to $|\mu |$ becoming
small, and comparable to the $SU(2)$ gaugino mass $M_2$). In the 
case of nearly degenerate and wino-like $\tz_1$ and $\tw_1$ states--
as occurs in generic AMSB models-- the mass limit 
on the light chargino extracted by
searches at LEP2 is that $m_{\tw_1}>91.9$ GeV\cite{lepw1lim}. Solutions 
with $m_{\tw_1}$ less than this limit occur in the shaded
region of the plot at low $m_{3/2}$, and so this region yields 
the low $m_{3/2}$ bound on HCAMSB parameter space around 
$m_{3/2}\sim 30$ TeV.\footnote{The LEP2 limit that $m_{H_{SM}}>114.4$ GeV
is also possibly constraining. However, we expect a theory error of 
$\sim \pm 3$ GeV on our calculated value of $m_h$. Since 
$m_h\agt 111$ GeV throughout the plot, we do not adopt any constraint
due to the Higgs mass.} 
The white-shaded regions all yield allowable superparticle mass spectra. 
The lowest value of $m_{\tg}$ which is accessible occurs at $m_{3/2}\sim 30$ TeV,
where $m_{\tg}\sim 730$ GeV. This value is far beyond any reasonable reach of the 
Fermilab Tevatron, so instead we focus in this paper on HCAMSB signatures at the 
CERN LHC.
For convenience, we
also show in Fig. \ref{fig:pspace} contours of $m_{\tg}$ and
$m_{\tu_L}=$ 1, 2 and 3 TeV, and also contours of 
$m_{\tst_1}=500$ and 1000 GeV, and $m_{\te_L}=350$ GeV.
The region with $m_{\te_L}\alt 350$ GeV may be accessible to probes
of direct slepton pair production at the LHC\cite{slepton}.

\subsection{$BF(b\to s\gamma )$ and $(g-2)_\mu$ in HCAMSB}

\begin{figure}[htbp]
\begin{center}
\includegraphics[angle=270,width=0.8\textwidth]{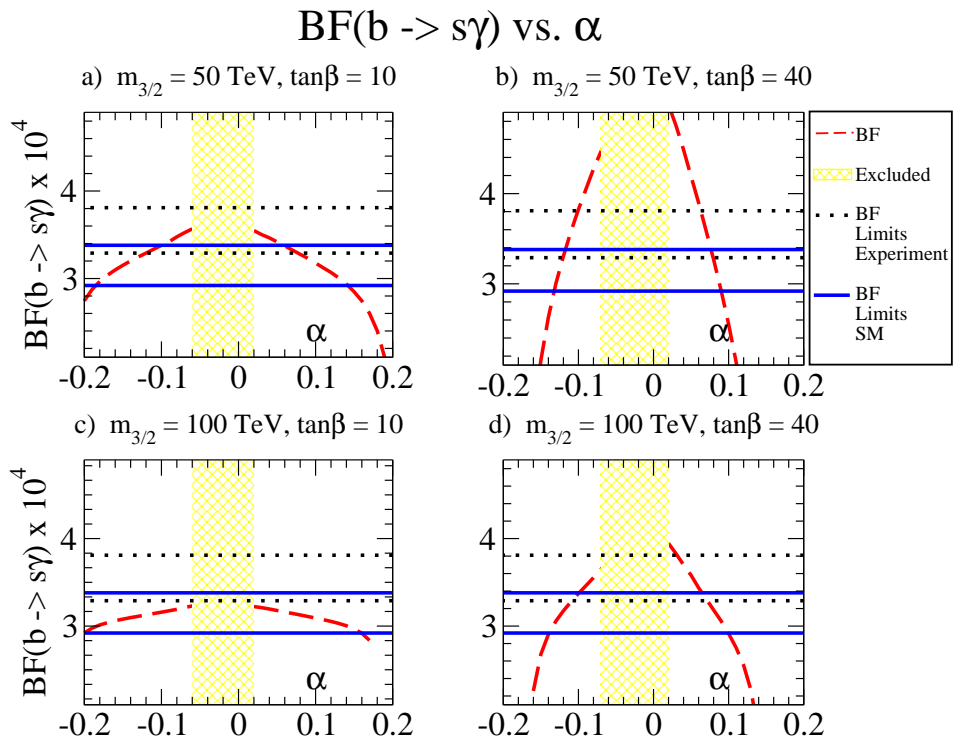}
\caption{Branching fraction for $b\to s\gamma$ versus $\alpha$ 
in the HCAMSB model
for $\mu >0$ and $(m_{3/2},\ tan\beta )=$ {\it a}). (50 TeV, 10), {\it b}). 
(50 TeV, 40), {\it c}). (100 TeV, 10)  and {\it d}). (100 TeV, 40). 
We also take $m_t=172.6$ GeV.
}
\label{fig:bsg}
\end{center}
\end{figure}

There also exist indirect limits on model parameter space from
comparing measured values of $BF(b\to s\gamma )$ and 
$\Delta a_\mu\equiv(g-2)_\mu$ against SUSY model predictions.
As an example, we show in Fig. \ref{fig:bsg} the branching fraction
for $BF(b\to s\gamma )$ in the HCAMSB model versus $\alpha$ 
for $m_{3/2}=50$ and 100 TeV, and for
$\tan\beta =10$ and 40 (calculated using the Isatools subroutine 
ISABSG\cite{isabsg}). We also show the region between the blue 
horizontal lines as the SM prediction ($BF(b\to s\gamma )_{SM}=(3.15\pm 0.23)\times 10^{-4}$ by a recent evaluation by Misiak\cite{misiak}), 
and the region between the 
black-dotted lines as the region allowed by experiment\cite{bsg_ex}.\footnote{
The branching fraction $BF(b\to s\gamma )$ has been measured by the
CLEO, Belle and BABAR collaborations; a combined analysis\cite{bsg_ex}
finds the branching fraction to be $BF(b\to s\gamma )=(3.55\pm                  
0.26)\times 10^{-4}$.}
The red-dashed curves show the HCAMSB prediction. 
We see that in each of the frames there exists some region of at least near
agreement with experiment. In frame {\it b}). with $m_{3/2}=50$ TeV
and $\tan\beta =40$, the low $\alpha$ region leads to too high of a
BF, while in frames {\it a})., {\it b}). and {\it d})., very high values
of $\alpha$ lead to too small a BF.

\begin{figure}[htbp]
\begin{center}
\includegraphics[angle=-90,width=0.8\textwidth]{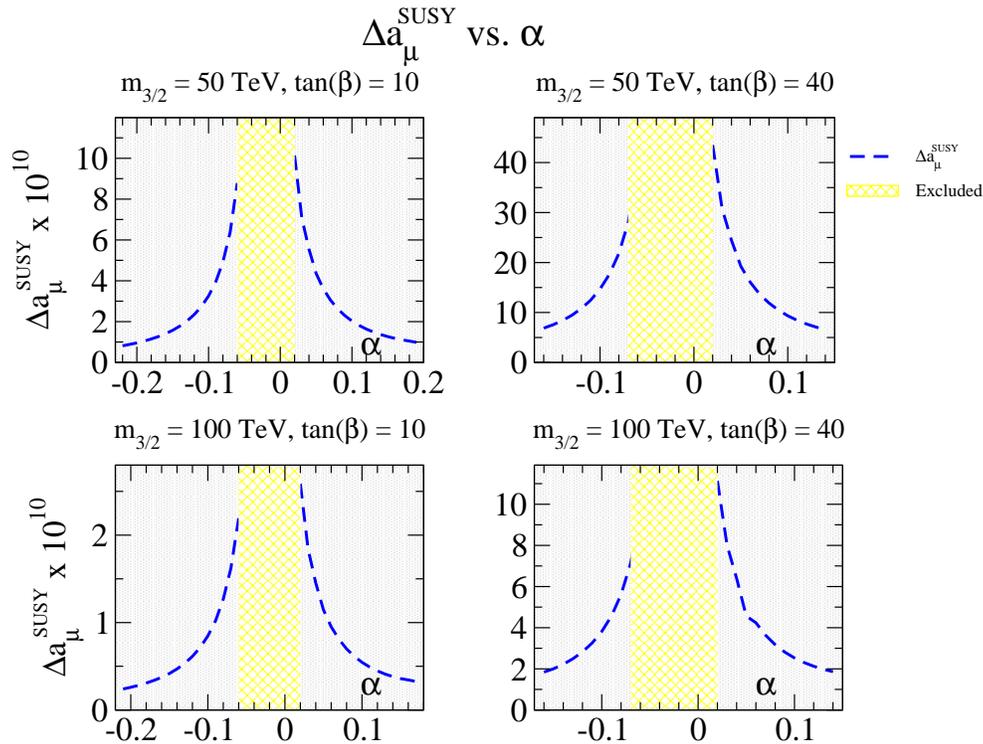}
\caption{SUSY contribution to $\Delta a_\mu$  versus $\alpha$ 
in the HCAMSB model
for $\mu >0$ and $(m_{3/2},\ tan\beta )=$ {\it a}). (50 TeV, 10), {\it b}). 
(50 TeV, 40), {\it c}). (100 TeV, 10)  and {\it d}). (100 TeV, 40). 
We also take $m_t=172.6$ GeV.
}
\label{fig:gm2}
\end{center}
\end{figure}

In Fig. \ref{fig:gm2}, we plot the SUSY contribution to $\Delta a_\mu$:
$\Delta a_\mu^{SUSY}$ (using ISAGM2 from Isatools\cite{isagm2}). 
The contribution is large when $\alpha$ is small;
in this case, rather light $\tmu_L$ and $\tnu_{\mu L}$ masses lead to
large deviations from the SM prediction. It is well-known that there
is a discrepancy between the SM predictions for $\Delta a_\mu$, where
$\tau$ decay data, used to estimate the hadronic vacuum polarization 
contribution to $\Delta a_\mu$, gives rough accord with the SM, while
use of $e^+e^-\to hadrons$ data at very low energy leads to a 
roughly $3\sigma$ discrepancy. 

Finally, we remark upon the relic density of dark matter in the HCAMSB model.
If thermal production of the lightest neutralino is assumed to give the 
dominant DM in the universe, then all over parameter space, the predicted
neutralino abundance $\Omega_{\tz_1}h^2$ is far below the WMAP 
measured value of $\Omega_{CDM}h^2\sim 0.11$. 
Some sample calculated values are listed in Table \ref{tab:cases}.
It has been suggested in Ref. \cite{moroi} that
production and decay of moduli fields or other processes can contribute
to the DM abundance. 
Decay of moduli fields in the early universe could then account for the
discrepancy between the measured DM abundance and the predicted thermal
abundance in HCAMSB models.
As an alternative, if the strong $CP$ problem is solved via the 
Peccei-Quinn mechanism, then a superfield containing the axion/axino multiplet
should occur. In this case, a mixture of axions\cite{axion} and axinos\cite{axino}, 
rather than wino-like neutralinos, could constitute the DM abundance\cite{mix}. 
In light of these two alternative DM mechanisms, we regard the 
HCAMSB parameter space as essentially unconstrained by the measured
abundance of DM in the universe.

\section{HCAMSB at the LHC}
\label{sec:lhc}

\subsection{Cross sections and branching fractions}

Across almost all of the HCAMSB model parameter space, we expect $\tw_1$ and $\tz_1$
to be wino-like, with $m_{\tw_1}\simeq m_{\tz_1}\sim {1\over 7.7} m_{\tg}$.
Thus, for the HCAMSB model, the dominant sparticle production cross sections 
at the LHC will consist of the $pp\to \tw_1^+\tw_1^- X$ and $pp\to \tw_1^\pm\tz_1 X$ 
reactions (as noted at the bottom of Table \ref{tab:cases}). 
These reactions will be very difficult-- if not impossible--
to observe, since they yield no energetic calorimeter deposition to serve as a trigger at 
LHC detectors.
Instead, sparticle detection prospects will have to rely on gluino and squark pair
production to yield observable collider events.

\begin{figure}[htbp]
\begin{center}
\includegraphics[angle=-90,width=0.7\textwidth]{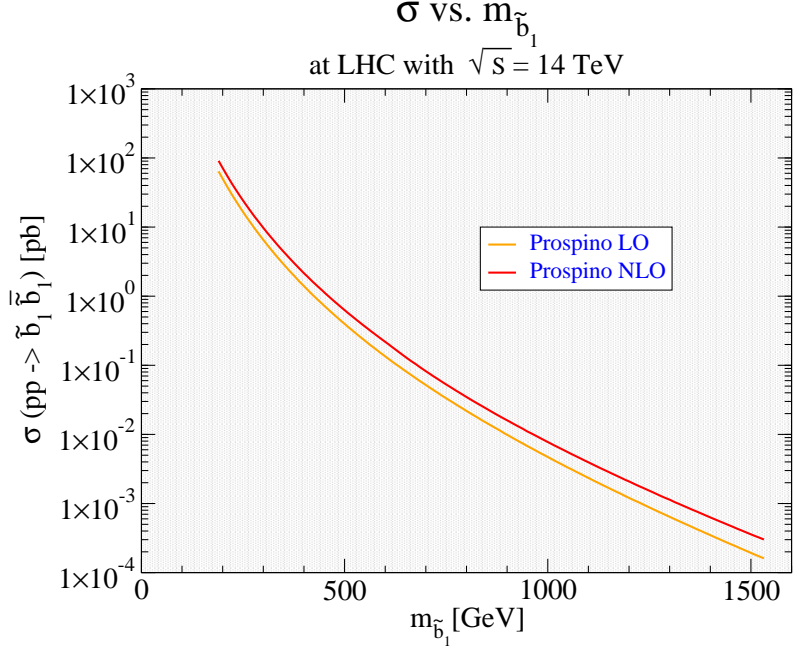}
\caption{Cross section for $pp\to \tb_1\bar{\tb}_1 X$ 
versus $m_{\tb_1}$ at LHC energy $\sqrt{s}=14$ TeV).
}
\label{fig:sigbb}
\end{center}
\end{figure}

At the lowest allowed values of $m_{3/2}\sim 30$ TeV, the gluino mass $m_{\tg}\sim 730$ GeV, 
and the combined $\tg\tg$, $\tg\tq$ and $\tq\tq$ pair production cross sections 
are of order $10^3-10^4$ fb\cite{wss}. At low 
$\alpha$ values, the value of $m_{\tg}$ is similar to $m_{\tq_L}$ and $m_{\tq_R}$ and 
all three of the above final states occur at similar rates.
In the high $\alpha$ regime of HCAMSB, the right squarks become quite heavy,
while third generation squark masses $\tst_1$ and $\tb_1$ become lighter.
In this case, $\tg\tg$ and $\tg\tu_L$ or $\tg\tc_L$ can occur at observable rates,
although the bulk of the strong production cross section can be dominated
by $\tst_1\bar{\tst}_1$ and $\tb_1\bar{\tb}_1$ production.
Since the $\tst_1$ and $\tb_1$ are dominantly left squarks at large $\alpha$, 
and are elements of a doublet, their masses are nearly equal, and their production
cross sections are similar. 
The direct $\tb_1\bar{\tb}_1$ production cross section is shown in 
Fig. \ref{fig:sigbb} for $pp$ collisions at $\sqrt{s}=14$ TeV\cite{prospino}.\footnote{
Initial LHC turn-on energy is expected to be around $\sqrt{s}=7-10$ TeV, 
with a gradual ramp-up towards $\sqrt{s}=14$ TeV. Cross sections
are of course model dependent, but generally we expect an increase in
cross sections of a factor of 2-4 in going from $\sqrt{s}=10$ TeV to 
$\sqrt{s}=14$ TeV. For instance, the $\sigma (pp\to t\bar{t}X)$ increases
by  a factor of 2.4 during this transition\cite{lhc}.}
The stop pair production rate is nearly identical since $m_{\tb_1}\simeq m_{\tst_1}$.

At low values of $\alpha$, since $m_{\tu_R,\tc_R}>m_{\tg}$, we get $\tu_R\to u\tg$
and $\tc_R\to c\tg$ adding to the gluino production rate. The 
$\tg$ decays mainly into $b\bar{\tb}_1+c.c.$ and $t\bar{\tst}_1+c.c.$
states, with a subdominant fraction of decays into other $q\tq_L$ pairs. 
As $\alpha$ increases, the right-squark masses increase, and ultimately decouple from 
the theory, while left-squark masses  increase slightly to values just above $m_{\tg}$. 
Thus, at high $\alpha$, the $\tg$ state decays purely into $b\bar{\tb}_1+c.c.$ and 
$t\bar{\tst}_1+c.c.$ pairs. 
We then expect that if strongly interacting sparticle states of the HCAMSB model are 
accessible to LHC searches, they should yield events with a high multiplicity of
$b$-quarks, $t$-quarks and $\tb_1$ and $\tst_1$ squarks, for all  values of $\alpha$.

\begin{figure}[htbp]
\begin{center}
\includegraphics[angle=-90,width=0.8\textwidth]{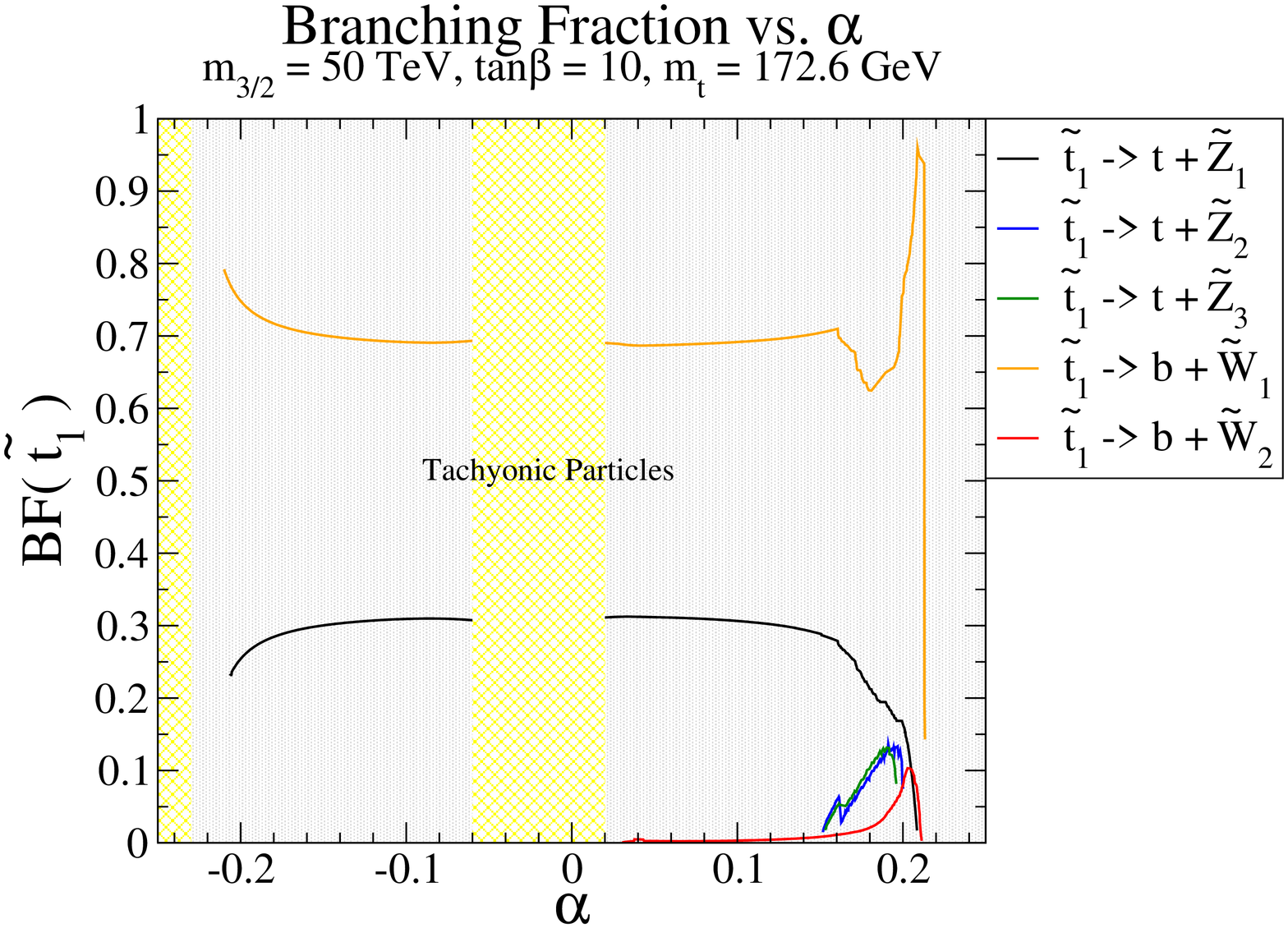}
\caption{Branching fraction of $\tst_1$ versus $\alpha$ for $m_{3/2}=50$ TeV,
$\tan\beta =10$ and $\mu >0$. 
}
\label{fig:bft1}
\end{center}
\end{figure}

\begin{figure}[htbp]
\begin{center}
\includegraphics[angle=-90,width=0.8\textwidth]{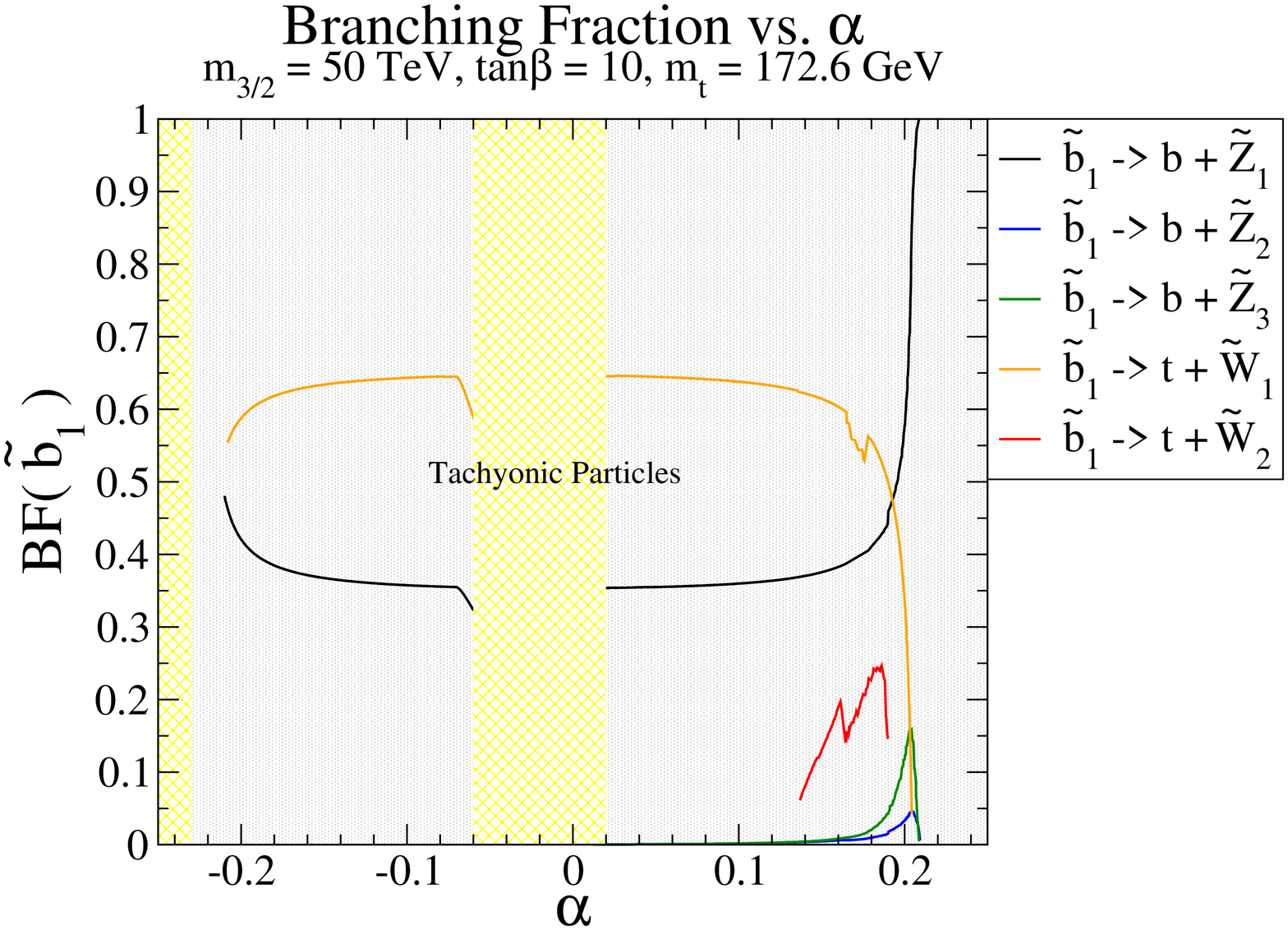}
\caption{Branching fraction of $\tb_1$ versus $\alpha$ for $m_{3/2}=50$ TeV,
$\tan\beta =10$ and $\mu >0$. 
}
\label{fig:bfb1}
\end{center}
\end{figure}

In Fig. \ref{fig:bft1} and Fig. \ref{fig:bfb1}, we show the $\tst_1$ and $\tb_1$ 
branching fractions versus $\alpha$ for $m_{3/2}=50$ TeV, $\tan\beta =10$ and $\mu >0$.
At low values of $\alpha$, we expect $\tst_1\to b\tw_1$ at $\sim 67\%$ and $\tst_1\to t\tz_1$
at $\sim 33\%$. Similarly, at low $\alpha$ we expect $\tb_1\to t\tw_1$ at $\sim 67\%$ and 
$\tb_1\to b\tz_1$ at $\sim 33\%$. 
As $|\alpha |$ increases, the value of $|\mu |$ decreases, until it becomes comparable to 
the gaugino mass $M_2$, and the $\tz_1$ state becomes mixed wino-higgsino. As $|\mu |$
decreases, so do the $\tz_2$, $\tz_3$ and $\tw_2$ eigenstates masses  
(while $m_{\tz_4}$ increases with mass $\sim M_1$ as it is nearly pure bino-like ). 
Thus, we see at large
$|\alpha |$, decay modes such as $\tb_1\to b\tz_2$, $b\tz_3$ and $t\tw_2$ turn-on,
leading to more complex cascade decays. Also, as $|\alpha |$ gets large, the modes
$\tst_1\to t\tz_2$, $t\tz_3$ and $b\tw_2$ become accessible (though never dominant).
Ultimately, as $|\alpha |$ increases even further, the values of $m_{\tst_1}$ and 
$m_{\tb_1}$ decrease, and the decay modes such as 
$\tst_1\to t\tz_3$, $t\tz_2$ and $t\tz_1$ all become kinematically suppressed. 
In fact, at the highest $\alpha$ values, the decay mode $\tst_1\to b\tw_1$
becomes kinematically dis-allowed, 
so that decays such as $\tst\to b\ell\nu \tz_1$ or $c\tz_1$ then dominate.

\begin{figure}[htbp]
\begin{center}
\includegraphics[angle=-90,width=0.7\textwidth]{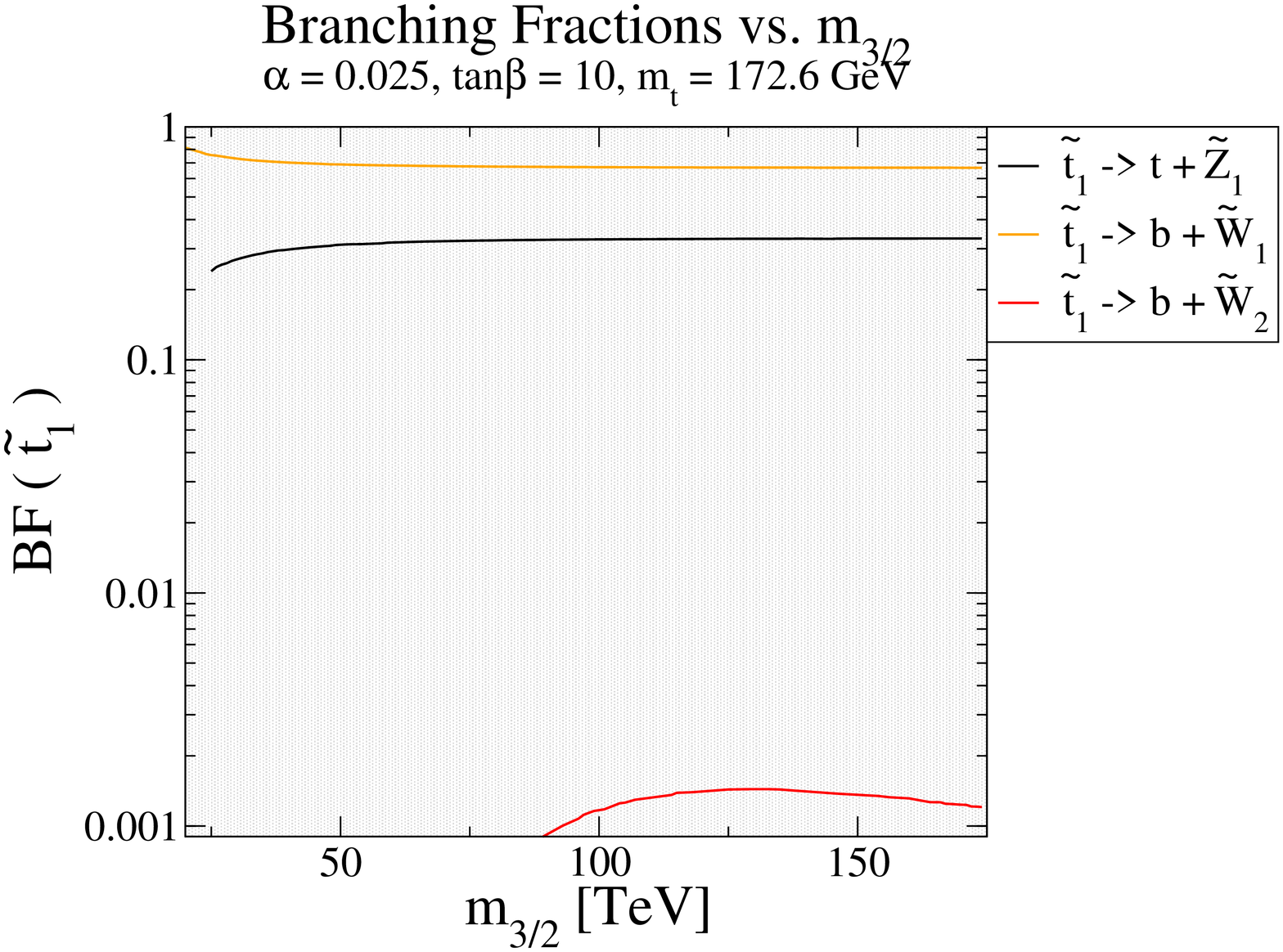}
\caption{Branching fraction of $\tst_1$ versus $m_{3/2}$ for $\alpha =0.025$,
$\tan\beta =10$ and $\mu >0$. 
}
\label{fig:bft1_g}
\end{center}
\end{figure}

\begin{figure}[htbp]
\begin{center}
\includegraphics[angle=-90,width=0.7\textwidth]{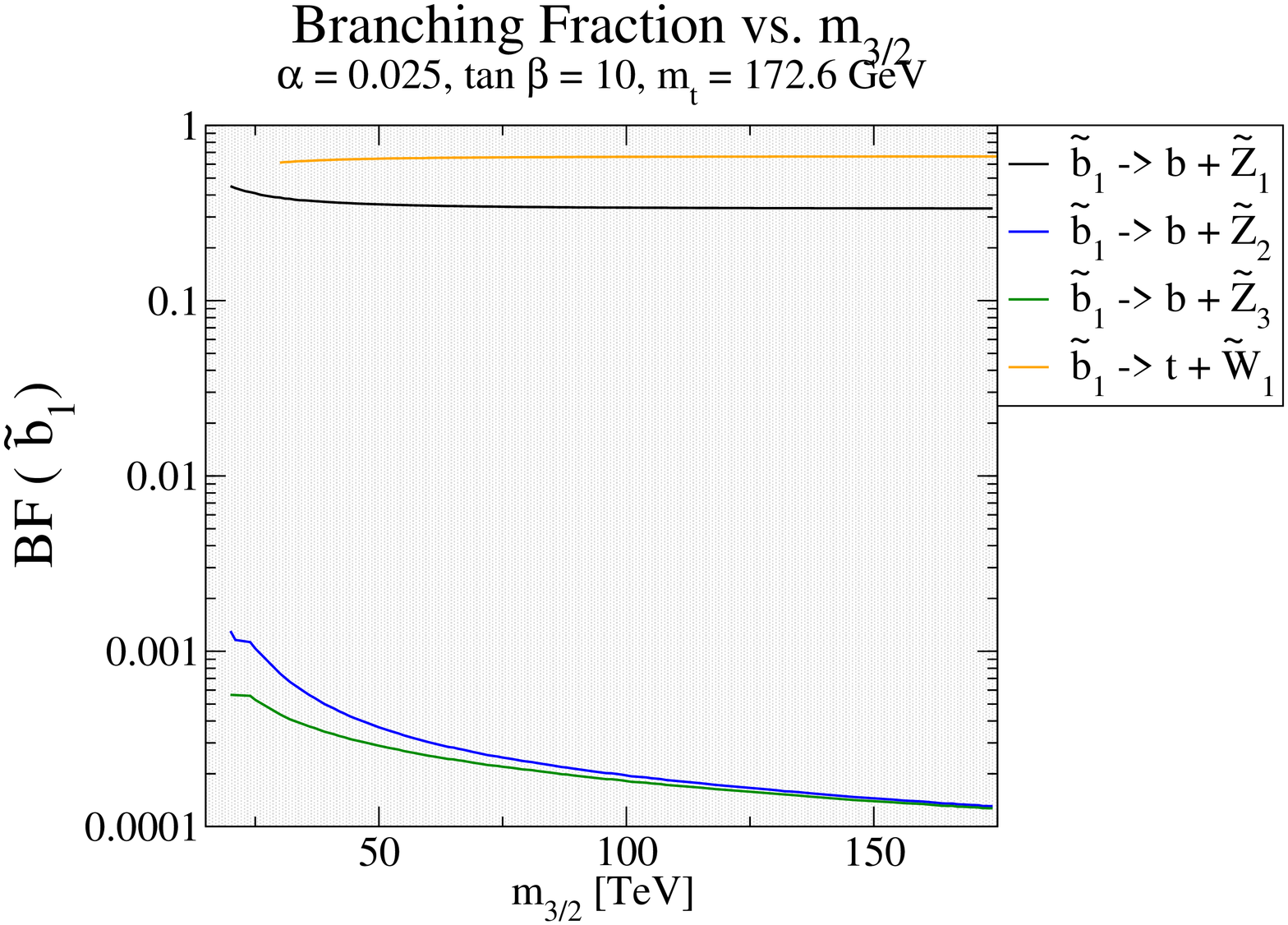}
\caption{Branching fraction of $\tb_1$ versus $m_{3/2}$ for $\alpha =0.025$,
$\tan\beta =10$ and $\mu >0$. 
}
\label{fig:bfb1_g}
\end{center}
\end{figure}

In Fig's \ref{fig:bft1_g} and \ref{fig:bfb1_g} we show the 
$\tst_1$ and $\tb_1$ branching fractions
versus $m_{3/2}$ for a fixed value of $\alpha =0.025$, $\tan\beta =10$ and $\mu >0$.
Here we see that $\tst_1\to b\tw_1$ and $t\tz_1$ dominates out to large $m_{3/2}$
values. This behavior persists also for high $\alpha$ values.
In the case of $\tb_1$, we see $\tb_1\to t\tw_1$ or $b\tz_1$ dominates over the
entire $m_{3/2}$ range as well.

Thus, in the HCAMSB model, we expect gluino and squark production events to 
cascade decay into third generation quarks and squarks. We then expect
HCAMSB collider events to contain a high multiplicity of $b$-jets, 
along with isolated leptons from $t\to bW$ decays,
and large $\eslt$ from escaping $\tz_1$ or $\nu$ states. Note as is usual in AMSB models
with $m_{\tw_1}\sim m_{\tz_1}$ that the $\tw_1$ is long-lived, and can fly distances of
order millimeters to centimeters before decaying via $\tw_1^+\to\pi^+\tz_1$ 
into a soft pion.
The presence of the highly ionizing chargino track, and its abrupt termination upon
chargino decay, is characteristic of models such as mAMSB and HCAMSB 
where the gaugino mass $M_2$ is far lighter than $M_1$ and $|\mu|$. 

\subsection{Characteristics of LHC collider events for cases HCAMSB1 and HCAMSB2}

We use Isajet 7.79\cite{isajet} for the simulation of signal and 
background events at the LHC. A toy detector simulation is employed with
calorimeter cell size
$\Delta\eta\times\Delta\phi=0.05\times 0.05$ and $-5<\eta<5$. The 
hadronic calorimeter (HCAL)
energy resolution is taken to be $80\%/\sqrt{E}+3\%$ for $|\eta|<2.6$ and
forward calorimeter (FCAL) is $100\%/\sqrt{E}+5\%$ for $|\eta|>2.6$. 
The electromagnetic (ECAL) energy resolution
is assumed to be $3\%/\sqrt{E}+0.5\%$. We use the UA1-like jet finding algorithm
GETJET with jet cone size $R=0.4$ and require that $E_T(jet)>50$ GeV and
$|\eta (jet)|<3.0$. Leptons are considered
isolated if they have $p_T(e\ or\ \mu)>20$ GeV and $|\eta|<2.5$ with 
visible activity within a cone of $\Delta R<0.2$ of
$\Sigma E_T^{cells}<5$ GeV. The strict isolation criterion helps reduce
multi-lepton backgrounds from heavy quark ($c\bar c$ and $b\bar{b}$) production.

We identify a hadronic cluster with $E_T>50$ GeV and $|\eta(j)|<1.5$
as a $b$-jet if it contains a $B$ hadron with $p_T(B)>15$ GeV and
$|\eta (B)|<3$ within a cone of $\Delta R<0.5$ about the jet axis. We
adopt a $b$-jet tagging efficiency of 60\%, and assume that
light quark and gluon jets can be mis-tagged as $b$-jets with a
probability $1/150$ for $E_T<100$ GeV, $1/50$ for $E_T>250$ GeV, 
with a linear interpolation for $100$ GeV$<E_T<$ 250 GeV\cite{xt}. 

We have generated 2M events each for cases HCAMSB1 and HCAMSB2 from Table \ref{tab:cases}.
In addition, we have generated background events using Isajet for
QCD jet production (jet-types include $g$, $u$, $d$, $s$, $c$ and $b$
quarks) over five $p_T$ ranges as shown in Table \ref{tab:bg}\cite{gabe}. 
Additional jets are generated via parton showering from the initial and final state
hard scattering subprocesses.
We have also generated backgrounds in the $W+jets$, $Z+jets$, 
$t\bar{t}(172.6)$ and $WW,\ WZ,\ ZZ$ channels at the rates shown in 
the same Table. The $W+jets$ and $Z+jets$ backgrounds
use exact matrix elements for one parton emission, but rely on the 
parton shower for subsequent emissions.

For our initial selection of signal events, we first require the following 
cuts labeled ${\bf C1}$:
\bi
\item $n(jets)\ge 4$,
\item $\eslt >max\ (100\ {\rm GeV},0.2M_{eff})$
\item  $E_T(j1,\ j2,\ j3,\ j4)>100,\ 50,\ 50,\ 50$ GeV,
\item  transverse sphericity $S_T>0.2$,
\ei
where $M_{eff}=\eslt +E_T(j1)+E_T(j2)+E_T(j3)+E_T(j4)$.

\begin{figure}[htbp]
\begin{center}
\includegraphics[width=0.8\textwidth]{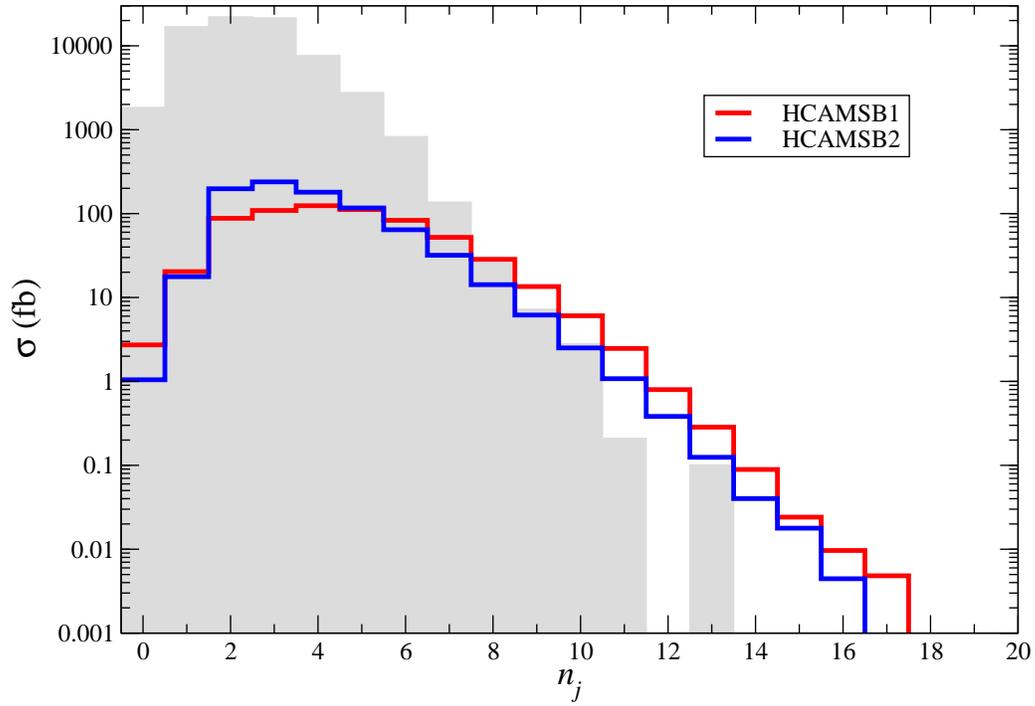}
\caption{Distribution in jet multiplicity in LHC collider events 
with $\sqrt{s}=14$ TeV from cases 
HCAMSB1 (red), HCAMSB2 (blue), and summed SM background (grey), after cuts set $C1$.
}
\label{fig:nj}
\end{center}
\end{figure}

In Fig. \ref{fig:nj}, we plot the resulting distribution in 
jet multiplicity (after relaxing the $n(jets)\ge 4$ requirement).
We see that the signal distributions for cases HCAMSB1 and HCAMSB2 are harder than the 
summed background histogram (gray), although signal doesn't exceed BG
until very high jet multiplicities around $n(jets)\sim 9$.
Thus, selecting signal events with $n(jets)\ge 2-4$ should be beneficial.

\begin{figure}[htbp]
\begin{center}
\includegraphics[width=0.8\textwidth]{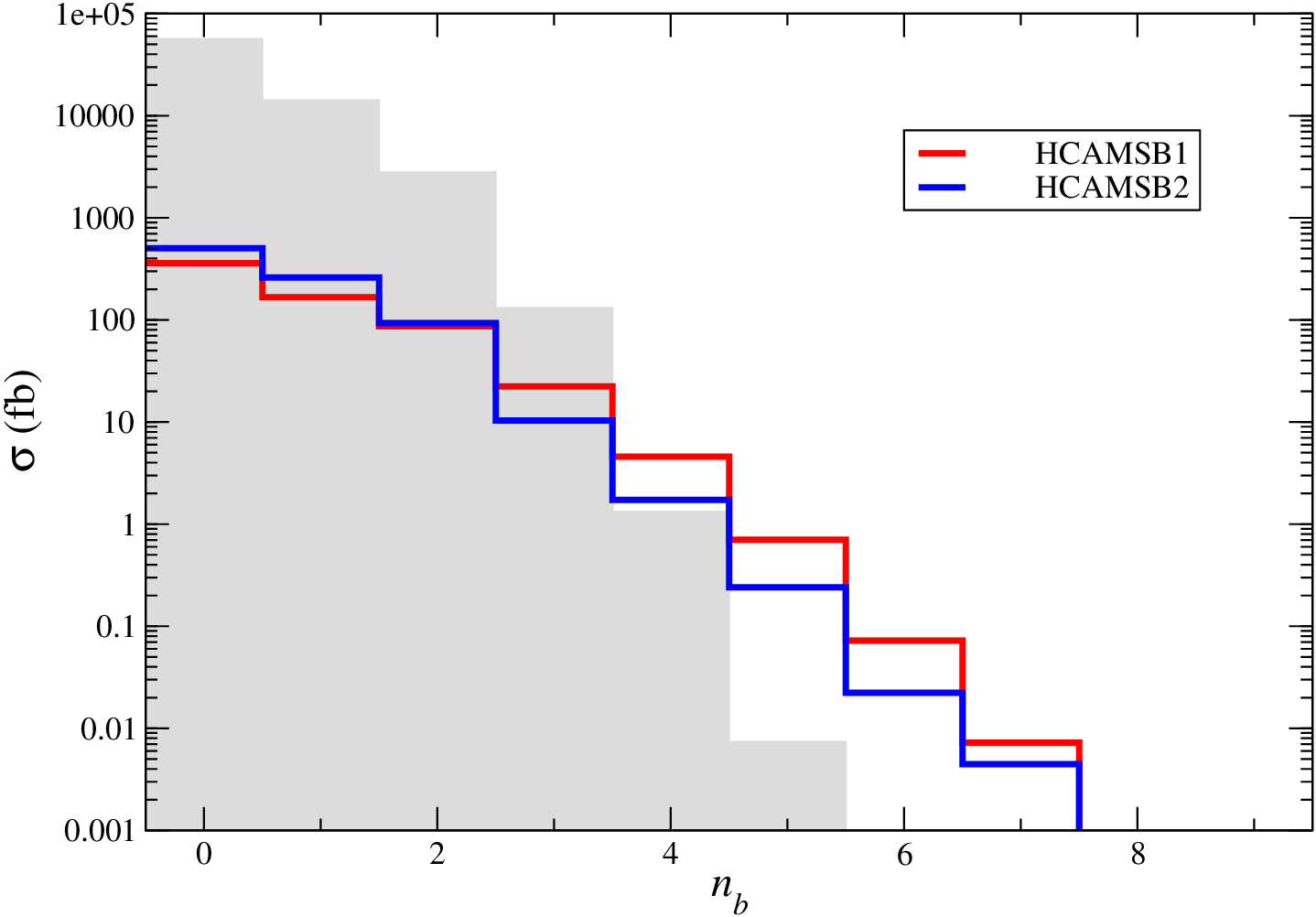}
\caption{Distribution in $b$-jet multiplicity in LHC collider events 
with $\sqrt{s}=14$ TeV from cases 
HCAMSB1 (red), HCAMSB2 (blue), and summed SM background (grey), after cuts set $C1$.
}
\label{fig:nbj}
\end{center}
\end{figure}

In Fig. \ref{fig:nbj}, we plot the distribution in $b$-jet multiplicity
from cases HCAMSB1 and HCAMSB2 against summed SM BG after cuts $C1$ 
(while again relaxing $n(jets)\ge 4$). As expected, the signal distributions 
are harder than the summed BG owing to the large number of $b$ and $t$
quarks produced in the HCAMSB cascade decay events. Signal
typically exceeds BG around $n(b-jets)\sim 5$.
Thus, requiring the presence of at least one identified $b$-jet will
aide in selecting HCAMSB signal over BG. 

\begin{figure}[htbp]
\begin{center}
\includegraphics[width=0.8\textwidth]{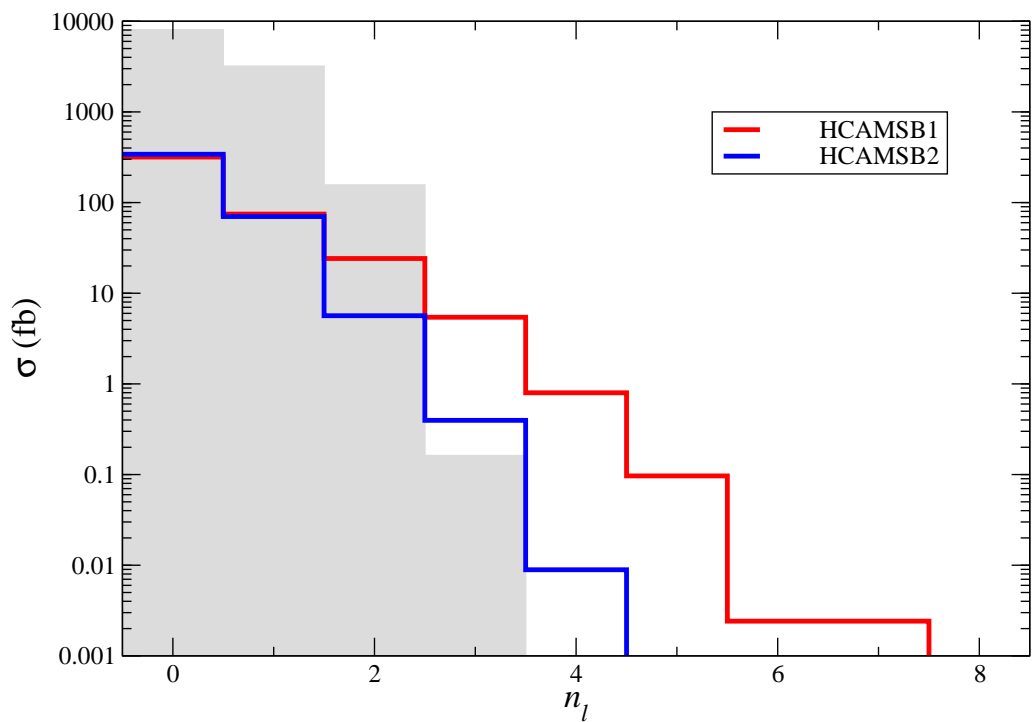}
\caption{Distribution in isolated lepton multiplicity in LHC collider events 
with $\sqrt{s}=14$ TeV from cases 
HCAMSB1 (red), HCAMSB2 (blue), and summed SM background (grey), after cuts set $C1$.
}
\label{fig:nl}
\end{center}
\end{figure}

In Fig. \ref{fig:nl}, we show the distribution in isolated 
lepton multiplicity after cuts $C1$. In this case, we see HCAMSB1, 
with its much lighter spectrum of sleptons, gives a much harder distribution 
in $n(leptons)$ than HCAMSB2. By $n(\ell )=3$, signal far exceeds BG,
especially for case HCAMSB1, where signal remains around 5 fb. 
This case should already be visible in early LHC SUSY searches with just a
few fb$^{-1}$ of integrated luminosity\cite{earlylhc}.

\begin{figure}[htbp]
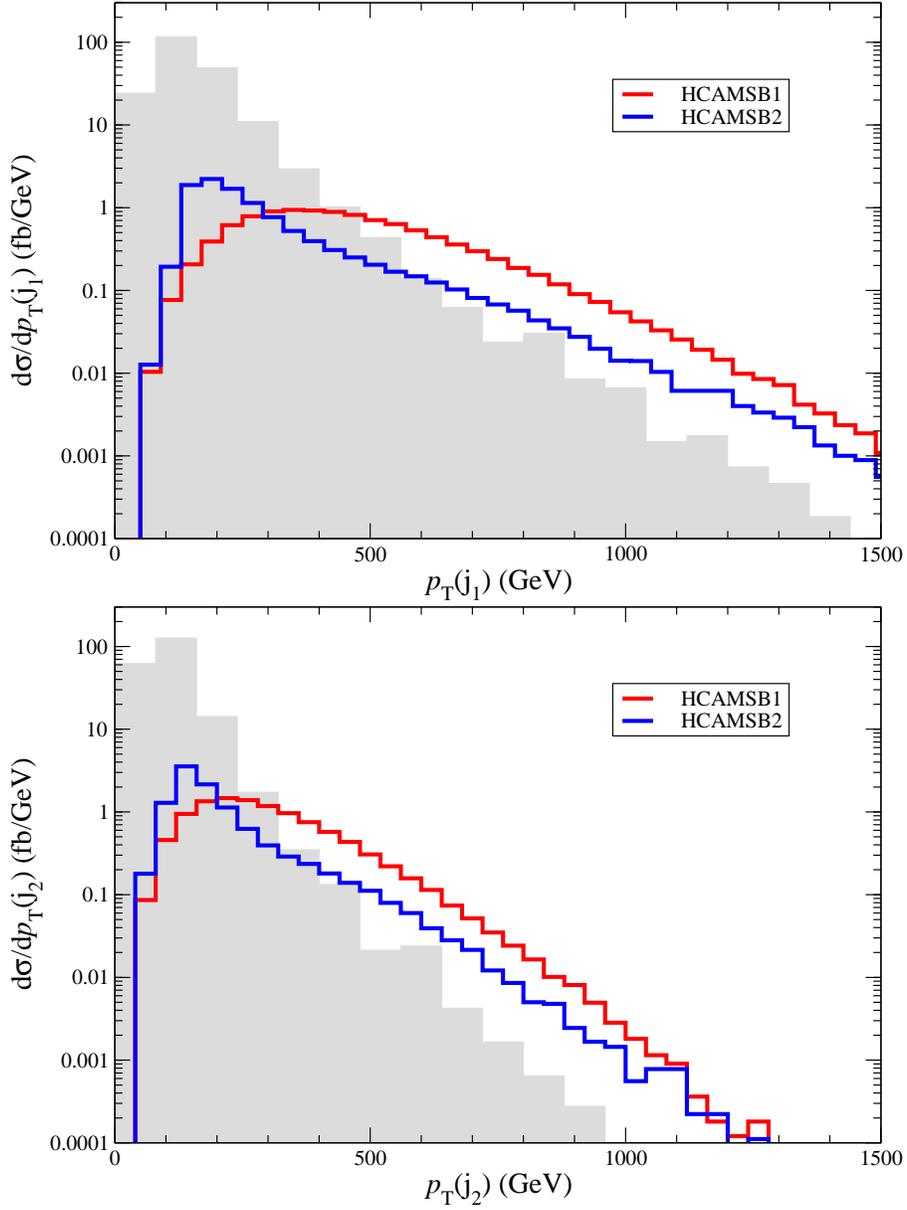

\begin{center}
\includegraphics[width=0.7\textwidth]{hcamsb_C1-etj1.eps}
\includegraphics[width=0.7\textwidth]{hcamsb_C1-etj2.eps}
\caption{Distribution in {\it a}). hardest jet $E_T$ 
and {\it b}). second hardest jet $E_T$ in LHC collider events 
with $\sqrt{s}=14$ TeV from cases 
HCAMSB1 (red), HCAMSB2 (blue), and summed SM background (grey), 
after cuts set $C1$.
}
\label{fig:etj1}
\end{center}
\end{figure}

Fig. \ref{fig:etj1} shows the distribution in $E_T$ of {\it a}). 
the hardest jet and {\it b}). the second hardest jet in
HCAMSB1 and HCAMSB2 events, along with SM BG after cuts $C1$ (but where the
hardest jet $E_T(j1)>100$ GeV cut is relaxed). Here, the case HCAMSB2 peaks
around $E_T(j1)\sim 150$ GeV, due mainly to $\tb_1\bar{\tb}_1$ production
followed by $\tb\to b\tz_1$ decay. Signal begins to exceed BG by around
450 GeV (HCAMSB1) or 550 GeV (HCAMSB2).

\begin{figure}[htbp]
\begin{center}
\includegraphics[width=0.8\textwidth]{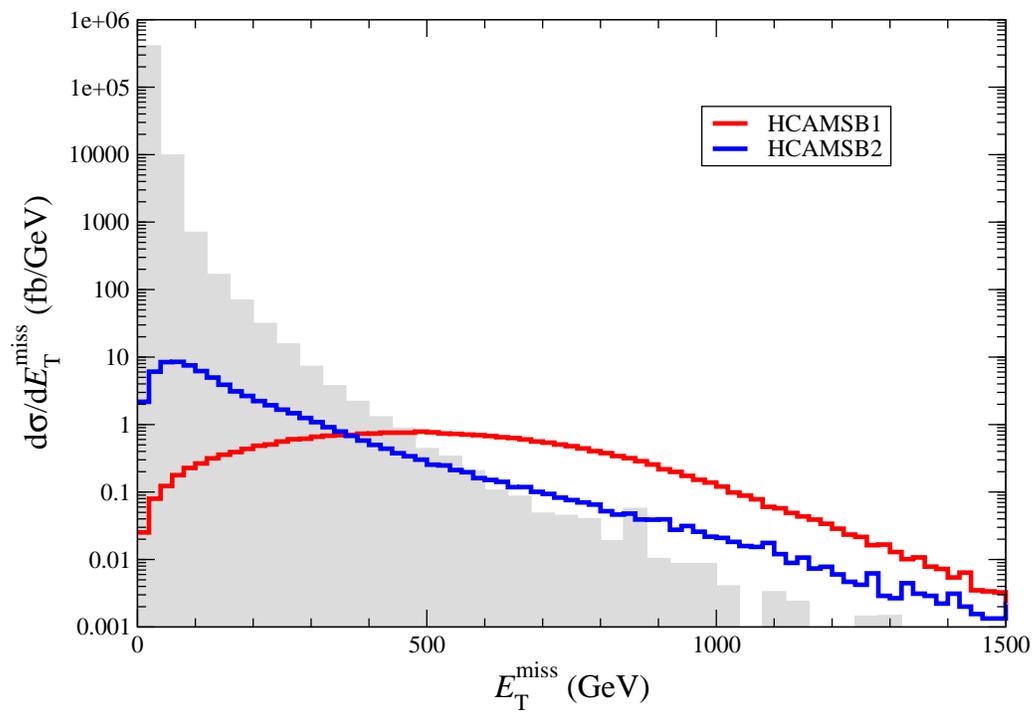}
\caption{Distribution in missing $E_T$ from LHC collider events 
with $\sqrt{s}=14$ TeV from cases 
HCAMSB1 (red), HCAMSB2 (blue), and summed SM background (grey), after cuts set $C1$.
}
\label{fig:etm}
\end{center}
\end{figure}

In Fig. \ref{fig:etm}, we show the missing $E_T$ distribution from
signal and BG events. The distribution from HCAMSB2, which is dominated by
relatively light 3rd generation squark production, is considerably 
softer than HCAMSB1, where production of TeV-scale squarks and gluinos
is dominant. Both cases exceed the summed BG for $\eslt\agt 500$ GeV.

\begin{figure}[htbp]
\begin{center}
\includegraphics[width=0.8\textwidth]{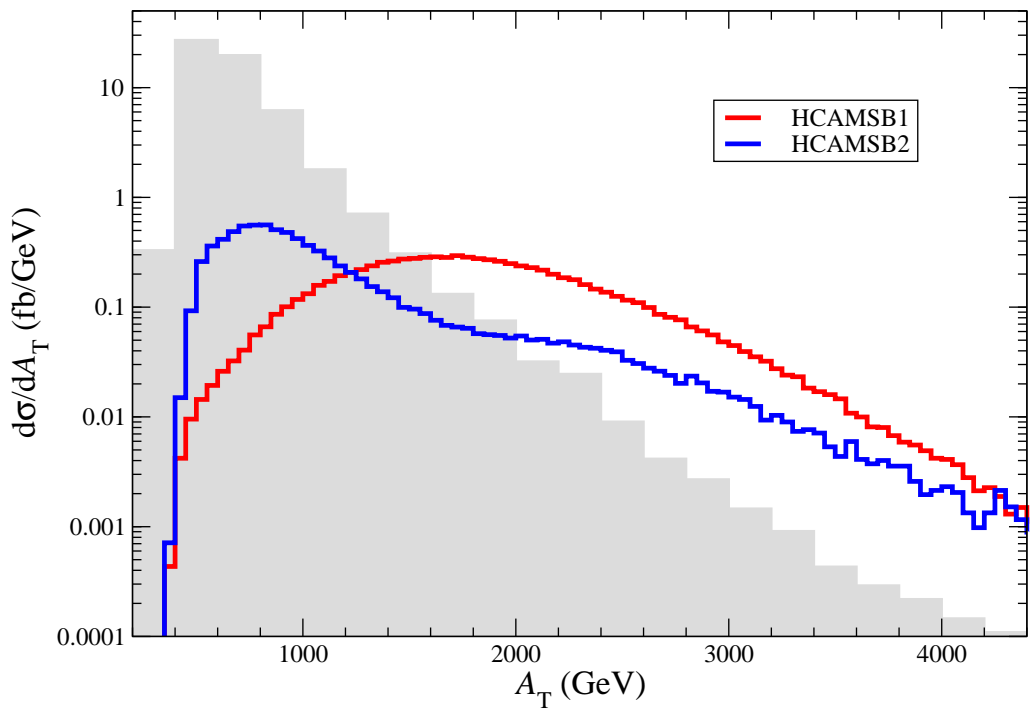}
\caption{Distribution in augmented effective mass $A_T$ from LHC collider events 
with $\sqrt{s}=14$ TeV from cases 
HCAMSB1 (red), HCAMSB2 (blue), and summed SM background (grey), after cuts set $C1$.
}
\label{fig:at}
\end{center}
\end{figure}

We  show in Fig. \ref{fig:at} the distribution in augmented effective mass 
$A_T=\eslt +\sum E_T(jets)+\sum E_T(isol.\ leptons)$. In this case, signal
point HCAMSB1 yields a rather smooth, hard distribution which emerges
from BG around $A_T\sim 1600$ GeV. Meanwhile, the $A_T$ distribution
from case HCAMSB2 actually resolves itself into two components:
a soft component peaks around $A_T\sim 750$ GeV, and is due to 3rd generation
squark pair production. The harder component, peaking around
$A_T\agt 2000$ GeV, occurs due to $\tg$ and $\tq_L$ production.

\subsubsection{LHC cascade decay events including HITs:
a smoking gun for AMSB models}

Of course, a distinctive property of models like HCAMSB (and also mAMSB) 
 with a wino-like $\tz_1$ state
is that the chargino is very long lived, as shown in Fig. \ref{fig:ctau}. Thus, 
once we have obtained cascade decay signal events in any of the multi-jet plus multi-lepton 
plus $\eslt$ channels, 
we may in addition
look for the presence of a highly-ionizing track (HIT) 
from the long-lived chargino. 
The presence of HITs in the SUSY collider events would be indictative of models
such as mAMSB or HCAMSB, where $M_2\ll M_1$ and $M_3$, so that the lightest
neutralino is a nearly pure wino state and where $m_{\tw_1}\simeq m_{\tz_1}$.

\subsubsection{Cascade decays including HITs plus $Z\to\ell\bar{\ell}$:
a smoking gun for HCAMSB?}

\begin{figure}[htbp]
\begin{center}
\includegraphics[width=0.8\textwidth]{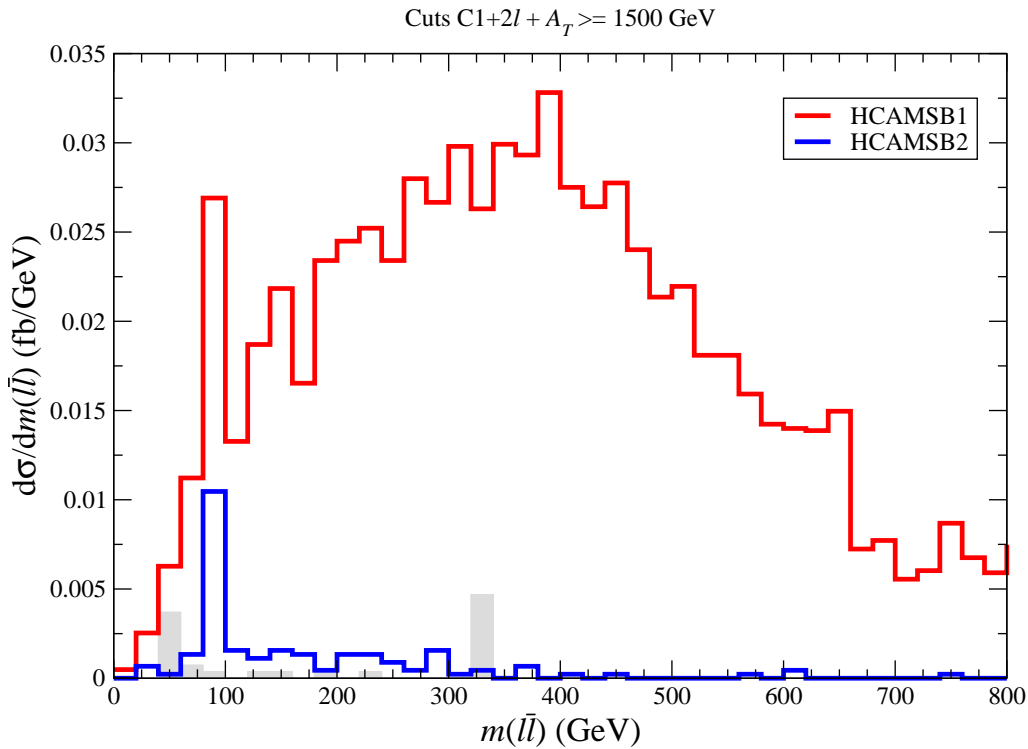}
\caption{Invariant mass distribution for same-flavor/opposite sign dileptons
from HCAMSB1 and HCAMSB2 after requiring cuts set $C1$ plus 
$A_T>1500$ GeV.
}
\label{fig:mll}
\end{center}
\end{figure}

Next we examine the distribution in $m(\ell^+\ell^- )$ for cascade decay events
containing two same-sign isolated dileptons (here, $\ell =e$ or $\mu$). 
This distribution has for long been touted
as being very useful as a starting point for reconstructing sparticle masses
in SUSY cascade decay events, because it may contain a kinematic mass edge from
$\tz_2\to\tell^\pm\ell^\mp$ or $\tz_2\to\ell^+\ell^-\tz_1$ decays. 
In the case of mAMSB models, such a mass edge may be present because
$\tz_2$ is bino-like and can decay into $\tell_R^\pm\ell^\mp$ at a high rate.
In the case of HCAMSB models, the $\tz_2$ state (and also the $\tz_3$ state) 
is expected
to be rather heavy and higgsino-like; it  decays mainly into two-body modes
such as $\tz_2\to \tw_1^\pm W^\mp$, $\tz_1 h$ and $\tz_2\to \tz_1 Z$. 
In particular, the later decay should
always be open (except when $\mu\to 0$ at the very highest $\alpha$ values) 
and can occur with branching fractions at the tens of percent level (see Table
\ref{tab:cases}). However, in mAMSB models, where $\tz_2$ is bino-like, its decay
to $\tz_1 Z$ is highly suppressed due to the structure of the
$\tz_1\tz_2 Z$ coupling (see Eq. 8.101 of Ref. \cite{wss}).
Thus, we would expect in HCAMSB models, 
instead of kinematic mass edges, a continuum distribution
in OS dilepton invariant mass, with a visible peak at $m(\ell^+\ell^- )\sim M_Z$.
In Fig. \ref{fig:mll}, we show the resulting distribution using cuts $C1$ plus
$A_T>1500$ GeV, to reduce SM BGs. As expected, the signal stands out well above
SM BG, but as a continuum, with a $Z$ peak. This distribution might serve as a 
``smoking gun'' LHC signature for HCAMSB models: we would expect-- in the case of 
HCAMSB models at the LHC-- cascade decay events with occasional HITs from the 
wino-like late-decaying charginos, but also with an OS dilepton spectrum
with a discernable $Z\to\ell^+\ell^-$ peak!
In mAMSB at high values of $\tan\beta$, mixing effects in the neutralino sector 
can also allow for some $Z\to\ell^+\ell^-$ cascade decay events.

\subsection{The reach of LHC for two HCAMSB model lines}

\begin{figure}[htbp]
\begin{center}
\includegraphics[angle=-90,width=0.7\textwidth]{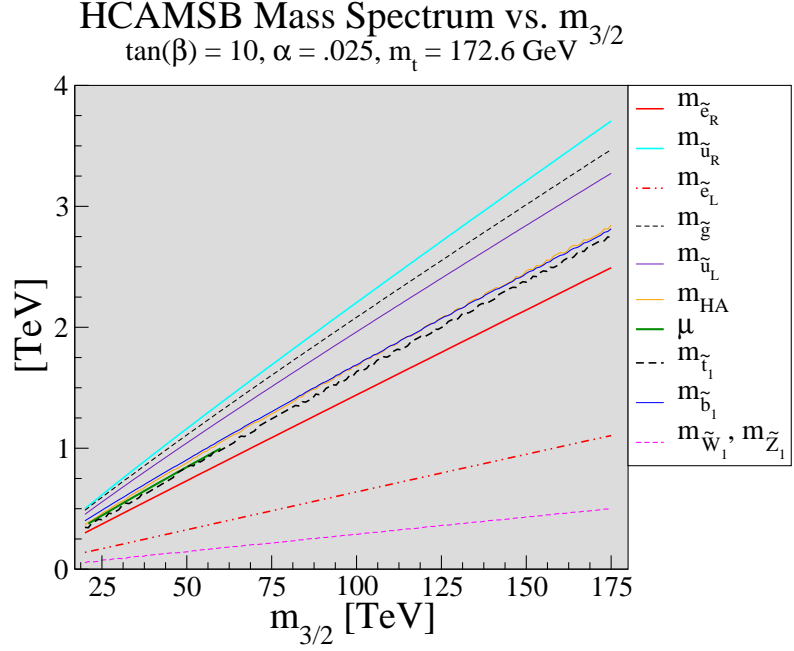}
\includegraphics[angle=-90,width=0.7\textwidth]{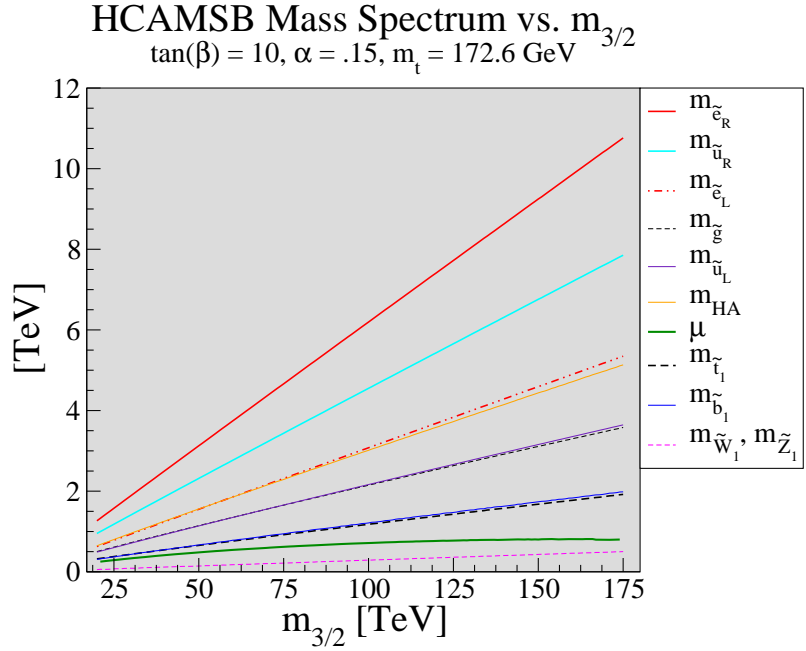}
\caption{Sparticle mass spectrum versus $m_{3/2}$ for HCAMSB model with 
{\it a}). $\alpha =0.025$ and {\it b}). $\alpha =0.15$
for $\tan\beta =10$, with $\mu >0$ and
$m_t=172.6$ GeV.
}
\label{fig:m32}
\end{center}
\end{figure}

We would next like to investigate the reach of the CERN LHC for SUSY in the
HCAMSB model. To this end, we will adopt two model lines. The first contains point HCAMSB1,
and so has $\alpha =0.025$, $\tan\beta =10$ and $\mu >0$. 
We will vary $m_{3/2}$ over the range
30 TeV to 200 TeV. For the second model line, we will take $\alpha =0.15$. 
We must take $\alpha$ somewhat 
lower than the HCAMSB2 point, since for $\alpha =0.195$, $m_{3/2}$ only extends up to 
about 60 TeV before hitting the EWSB-disallowed region (from Fig. \ref{fig:pspace}).
The sparticle mass spectra versus $m_{3/2}$ is shown for each of the two model lines
in Fig. \ref{fig:m32}.

Motivated by the previous signal and background distributions, 
we will require the following cuts $C2$\cite{bbkt}:
\begin{itemize}
\item $n(jets)\ge 2$
\item $S_T>0.2$
\item $n(b-jets)\ge 1$
\item $E_T(j1),\ E_T(j2),\ \eslt >E_T^c$,
\end{itemize} 
where $E_T^c$ can be variable. Parameter space points with lower sparticle masses
will benefit from lower choices of $E_T^c$, while points with heavier sparticle
masses, lower cross sections but higher energy release per event, will benefit from
higher choices of $E_T^c$.
In addition, in the zero-leptons channel we require 
$30^\circ <\Delta\phi (\vec{E}_T^{miss},{\vec{E}_T}(j_c))<90^\circ$ between the $\vec{E}_T^{miss}$ 
and the nearest jet in transverse opening angle. 
For all isolated leptons $\ell$, we require $p_T(\ell )>20$ GeV,
and for events with a single isolated lepton, we require the transverse mass 
$M_T(\ell,\eslt )\ge 100$ GeV to reject background events from $W\to\ell\nu_\ell$ production.
We separate the signal event channels according to the multiplicity of isolated
leptons: the $0\ell$, $1\ell$, same-sign (SS) and opposite-sign (OS) dilepton, 
and $3\ell$ channels. We do not here require ``same flavor'' on the SS or OS dilepton events.

The resultant cross sections after cuts $C2$ for SM backgrounds along with 
signal points HCAMSB1 and HCAMSB2 are listed in Table \ref{tab:bg} for $E_T^c=100$ GeV.
For each BG channel, we have generated $\sim 2$ million simulated events. 
With the hard cuts $C2$, we are unable to pick up BG cross sections in
some of the multi-lepton channels. 
We will consider a signal to be observable at an assumed value of integrated luminosity 
if {\it i})~the signal to
background ratio, $S/B \ge 0.2$, {\it ii})~the signal has a minimum of
five events, and {\it iii})~the signal satifies a statistical criterion
$S \ge 5\sqrt{B}$ (a $5\sigma$ effect).
%
\begin{table}
\begin{center}
\begin{tabular}{lccccc}
\hline
process & $0\ell$ & $1\ell$ & $OS$ & $SS$ & $3\ell$  \\
\hline
QCD($p_T$: 0.05-0.10 TeV) & -- & -- & -- & -- & -- \\
QCD($p_T$: 0.10-0.20 TeV) & -- & -- & -- & -- & -- \\
QCD($p_T$: 0.20-0.40 TeV) & 73.5 & -- & -- & -- & -- \\
QCD($p_T$: 0.40-1.00 TeV) & 42.6 & 26.5 & 37.3 & -- & -- \\
QCD($p_T$: 1.00-2.40 TeV) & 0.8 & 0.6 & 0.3 & 0.015 & -- \\
$t\bar{t}$ & 1253.2 & 341.2 & 224.9 & $ 0.25 $ & $ 0.25 $ \\
$W+jets; W\to e,\mu,\tau$ & 60.6 & 5.6  & 2.8 & $ -- $ & $ -- $ \\
$Z+jets; Z\to \tau\bar{\tau},\ \nu s$ & 61.4 & 0.0 & 0.77 & $ -- $ & $ -- $ \\
$WW,ZZ,WZ$ & 0.11 & --  & $ -- $ & $ -- $ & $ -- $ \\
\hline
$summed\ SM\ BG$ & 1492.3 & 374.1 & 266.1 & $ 0.26 $ & $ 0.25 $ \\
\hline
HCAMSB1 & 100.1 & 53.2 & 13.1 & 2.4 & 3.3 \\
HCAMSB2 & 223.5 & 58.7 & 4.6 & 1.7 & 0.35 \\
\hline
\end{tabular}
\caption{Estimated SM background cross sections (plus two HCAMSB benchmark 
points) in fb for various multi-lepton
plus jets $+\eslt$ topologies after cuts C2 with $E_T^c=100$ GeV.
}
\label{tab:bg}
\end{center}
\end{table}

\begin{figure}[htbp]
\begin{center}
\includegraphics[angle=-90,width=0.8\textwidth]{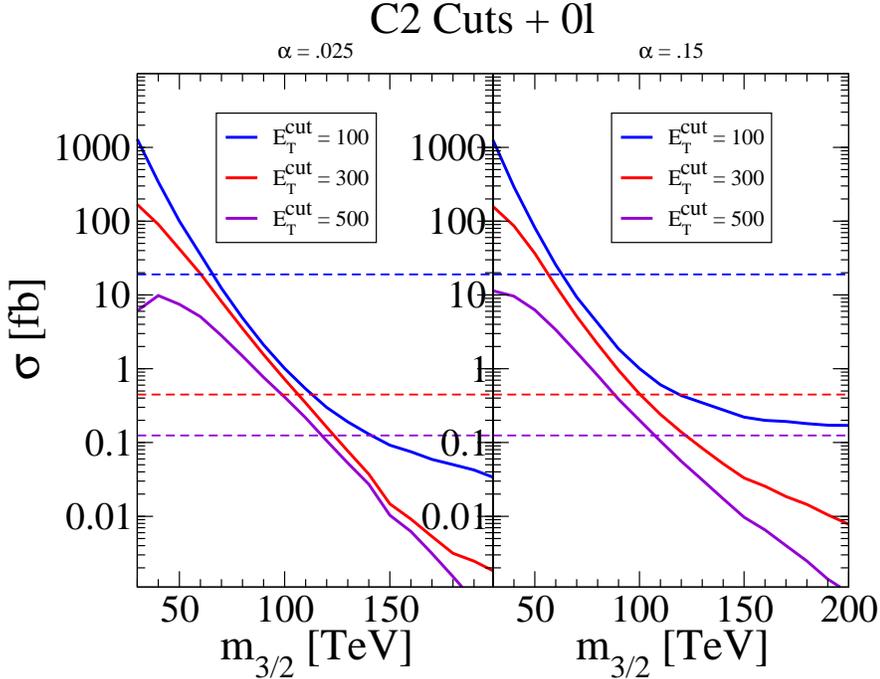}
\caption{Cross section for multi-jet plus $\eslt$
events with $n(\ell) =0$ at the LHC after cuts $C2$ listed in 
the text with $E_T^c=100,\ 300$ and 500 GeV, versus $m_{3/2}$, 
from HCAMSB model points 
with $\tan\beta =10$, $\mu >0$ and {\it a}). $\alpha =0.025$ and
{\it b}). $\alpha =0.15$. We also list the 100 fb$^{-1}$ $5\sigma$ 
reach with the horizontal lines.  
}
\label{fig:sig_0l}
\end{center}
\end{figure}

Using the above criteria, the 100 fb$^{-1}$ reach of the LHC can be computed
for each signal channel. 
In Fig. \ref{fig:sig_0l}, we show the signal rates versus $m_{3/2}$ 
for each of the two model lines for $E_T^c=100$, 300 and 500 GeV.
The $5\sigma$/ 5 event, 100 fb$^{-1}$ reach is denoted by the 
horizontal lines for each $E_T^c$ value.
We see the LHC reach in the $0\ell$ channel extends to
$m_{3/2}\sim 65,\ 105$ and 115 TeV for $E_T^c=100,\ 300$ and 500 GeV,
respectively, for the $\alpha =0.025$ case. This corresponds to a reach in
$m_{\tg}$ of 1.4, 2.2 and 2.4 TeV. The $\alpha=0.15$ model, shown in frame
{\it b})., exhibits a 100 fb$^{-1}$ reach of $m_{3/2}=60,\ 100$ and 105 TeV
for each $E_T^c$ value, corresponding to a reach in $m_{\tg}$ of 
1.3, 2.1 and 2.2 TeV, respectively. The reach for the high $\alpha$ model
line is somewhat lower than the low $\alpha$ model line since many of the
squark masses increase severely with $\alpha$, and no longer contribute
to the signal events.

\begin{figure}[htbp]
\begin{center}
\includegraphics[angle=-90,width=0.8\textwidth]{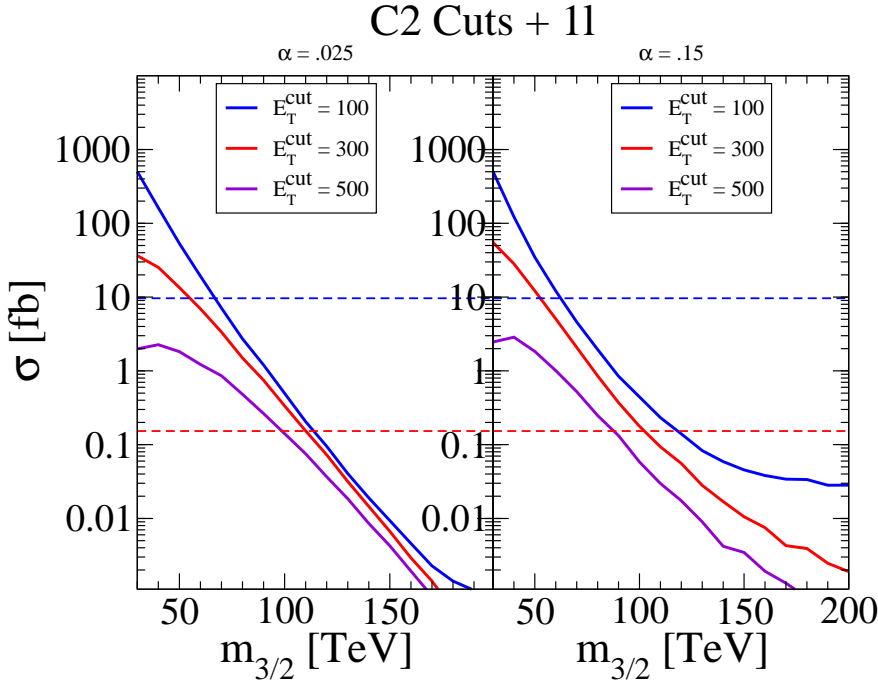}
\caption{Cross section for multi-jet plus $\eslt$
events with $n(\ell) =1$ at the LHC after cuts $C2$ listed in 
the text with $E_T^c=100,\ 300$ and 500 GeV, versus $m_{3/2}$, 
from HCAMSB model points 
with $\tan\beta =10$, $\mu >0$ and {\it a}). $\alpha =0.025$ and
{\it b}). $\alpha =0.15$. We also list the 100 fb$^{-1}$ $5\sigma$ 
reach with the horizontal lines.  
}
\label{fig:sig_1l}
\end{center}
\end{figure}

\begin{figure}[htbp]
\begin{center}
\includegraphics[angle=-90,width=0.8\textwidth]{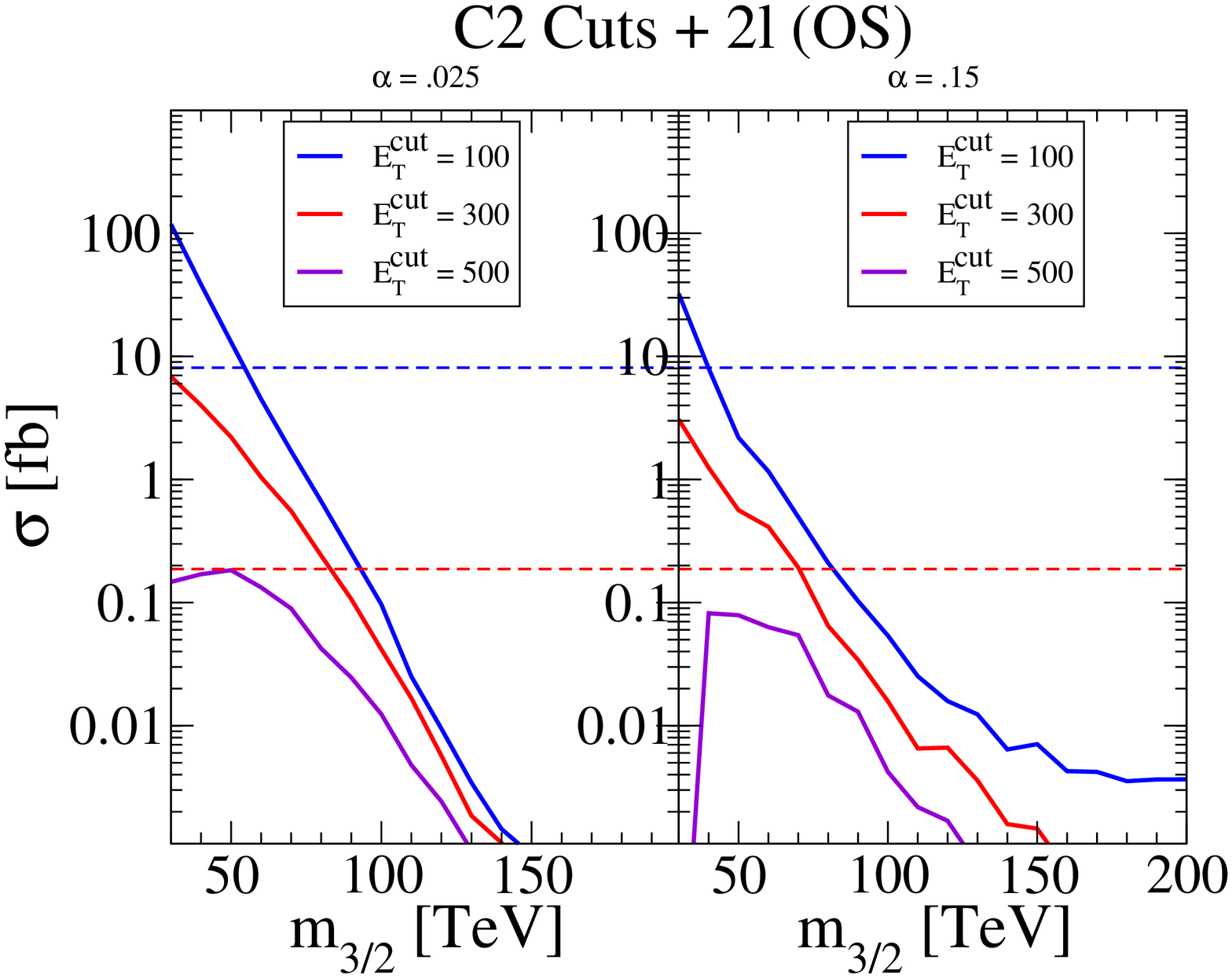}
\caption{Cross section for multi-jet plus $\eslt$
events with $n(\ell) =2$ (OS) at the LHC after cuts $C2$ listed in 
the text with $E_T^c=100,\ 300$ and 500 GeV, versus $m_{3/2}$, 
from HCAMSB model points 
with $\tan\beta =10$, $\mu >0$ and {\it a}). $\alpha =0.025$ and
{\it b}). $\alpha =0.15$. We also list the 100 fb$^{-1}$ $5\sigma$ 
reach with the horizontal lines.  
}
\label{fig:sig_OS}
\end{center}
\end{figure}

\begin{figure}[htbp]
\begin{center}
\includegraphics[angle=-90,width=0.8\textwidth]{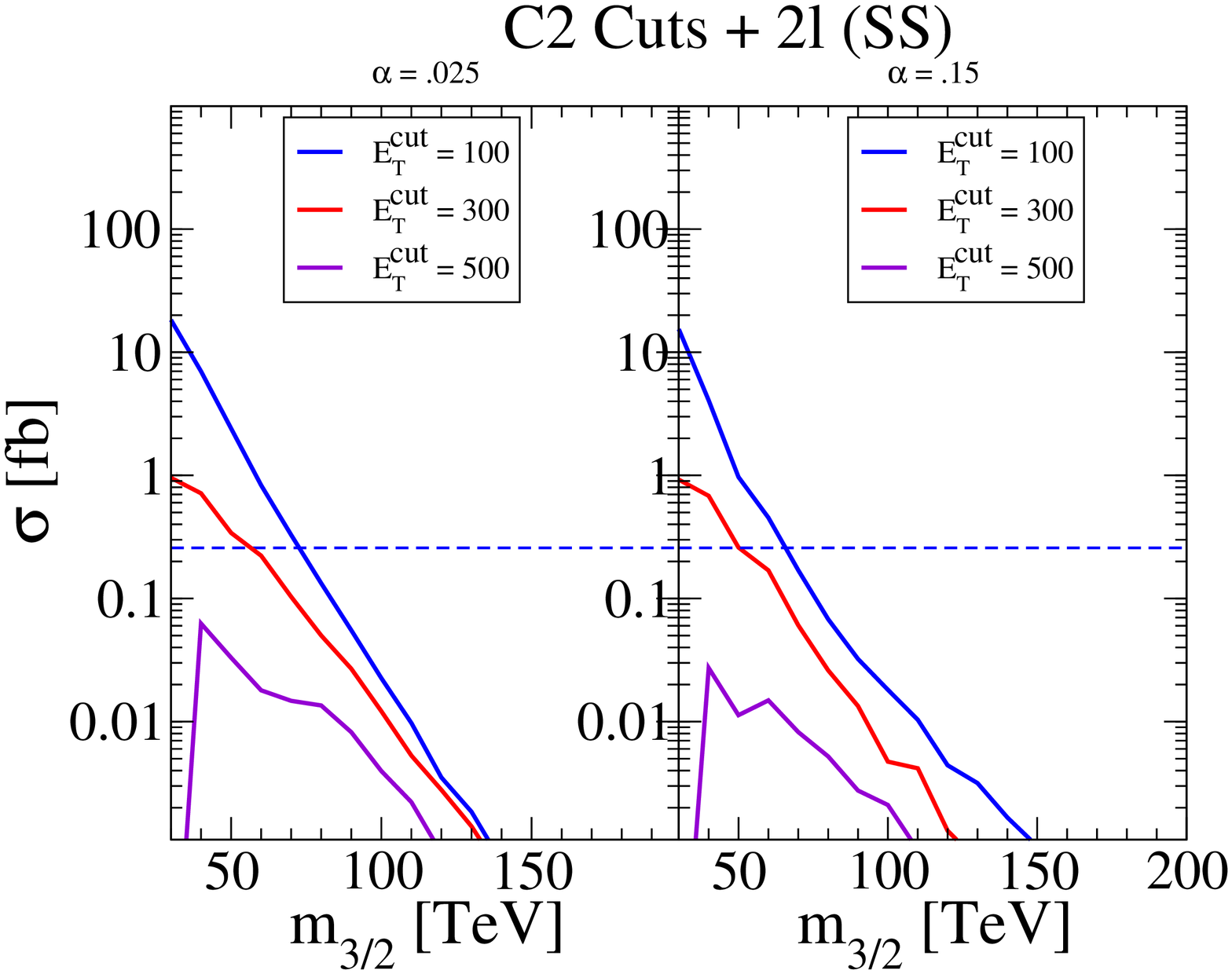}
\caption{Cross section for multi-jet plus $\eslt$
events with $n(\ell) =2$ (SS) at the LHC after cuts $C2$ listed in 
the text with $E_T^c=100,\ 300$ and 500 GeV, versus $m_{3/2}$, 
from HCAMSB model points 
with $\tan\beta =10$, $\mu >0$ and {\it a}). $\alpha =0.025$ and
{\it b}). $\alpha =0.15$. We also list the 100 fb$^{-1}$ $5\sigma$ 
reach with the horizontal lines.  
}
\label{fig:sig_SS}
\end{center}
\end{figure}

\begin{figure}[htbp]
\begin{center}
\includegraphics[angle=-90,width=0.8\textwidth]{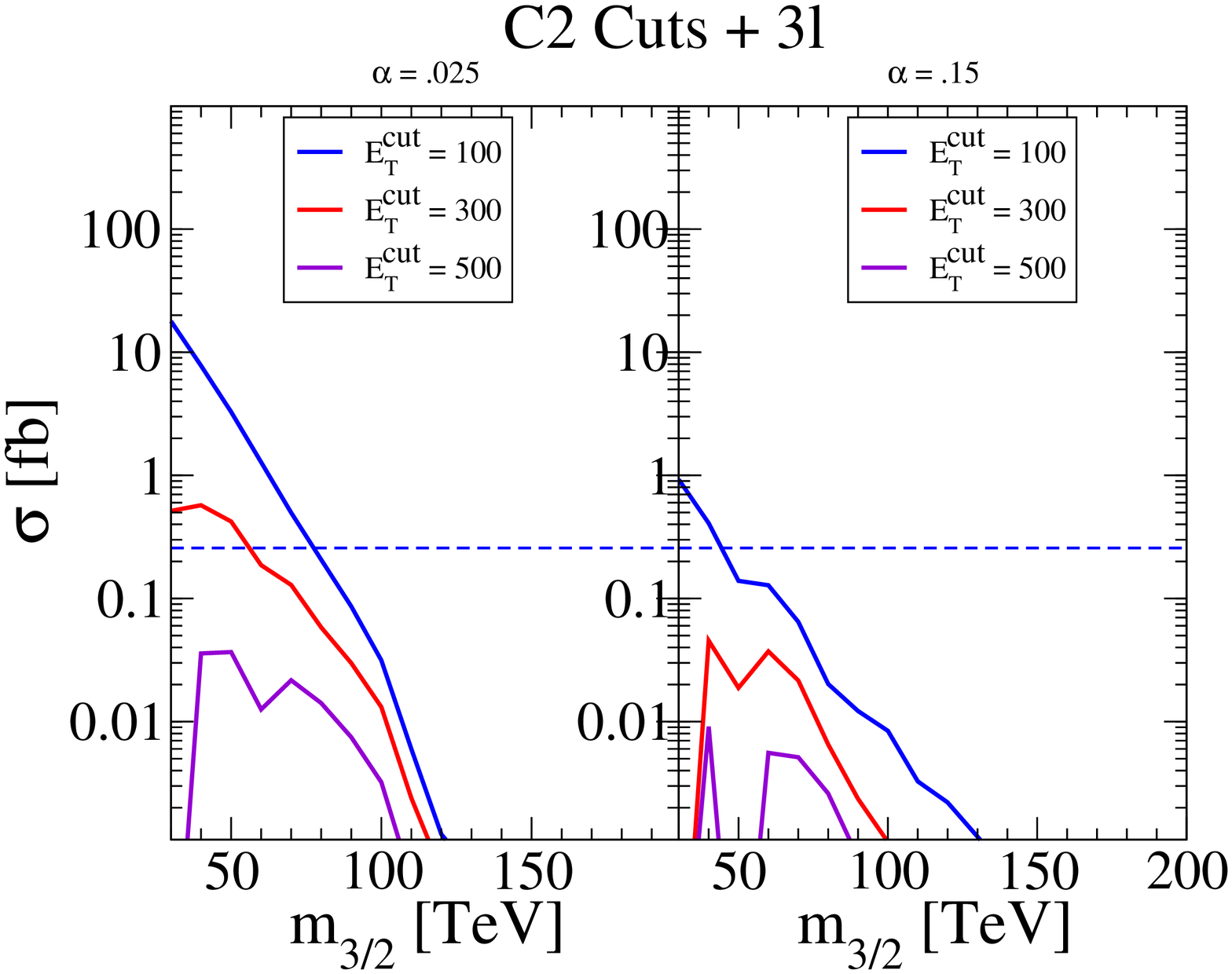}
\caption{Cross section for multi-jet plus $\eslt$
events with $n(\ell) =3$ at the LHC after cuts $C2$ listed in 
the text with $E_T^c=100,\ 300$ and 500 GeV, versus $m_{3/2}$, 
from HCAMSB model points 
with $\tan\beta =10$, $\mu >0$ and {\it a}). $\alpha =0.025$ and
{\it b}). $\alpha =0.15$. We also list the 100 fb$^{-1}$ $5\sigma$ 
reach with the horizontal lines.  
}
\label{fig:sig_3l}
\end{center}
\end{figure}

In Fig's \ref{fig:sig_1l}-\ref{fig:sig_3l}, 
we show the corresponding 100 fb$^{-1}$ reach of LHC for
the two HCAMSB model lines in the $1\ell$, $OS$, $SS$ and $3\ell$ channels.
We do not exhibit a $5\sigma$ horizontal line for those cases where 
we generate no surviving background events.
The reach in terms of $m_{3/2}$ for all channels is summarized in Table \ref{tab:reach}.
For a given $E_T^c$ value and signal channel, the upper entry corresponds to
the $\alpha =0.025$ model line, while the lower entry corresponds to the 
$\alpha =0.15$ model line. By examining Table \ref{tab:reach}, we see that the
maximal reach of LHC with 100 fb$^{-1}$ for the $\alpha =0.025$ model line
occurs in the $3\ell$ channel for $E_T^c=100$ GeV, with $m_{3/2}\sim 80$ GeV
being probed. However, a higher reach can be obtained by going to harder
cuts with $E_T^c=500$ GeV in the $0\ell$ channel, where the 
reach extends to $m_{3/2}\sim 115$ GeV, corresponding to a reach in 
$m_{\tg}$ of $\sim 2.4$ TeV. The maximal LHC reach for the $\alpha =0.15$ model
line with $E_T^c=100$ GeV occurs in the $1\ell$ and SS dilepton channels, with 
$m_{3/2}=65$ GeV being probed. 
The best reach for $\alpha =0.15$ can be obtained
using $E_T^c=500$ GeV in the $0\ell$ channel, where $m_{3/2}\sim 105$ TeV can be probed,
corresponding to a reach in $m_{\tg}$ of about 2.2 TeV.
%

%
\begin{table}
\begin{center}
\begin{tabular}{lccccc}
\hline
$E_T^c$ (GeV) & $0\ell$ & $1\ell$ & $OS$ & $SS$ & $3\ell$  \\
\hline
100 & $65/60$ & $65/65$ &  $55/ 40$ & $70/ 65$ & $80/ 45$ \\
300 & $105/100$ & $110/105$ &  $85/70$ & $-/-$ & $-/-$ \\
500 & $115/ 105$ & $-/-$ &  $-/ -$ & $-/ -$ & $-/ -$ \\
\hline
\end{tabular}
\caption{Estimated reach of 100 fb$^{-1}$ LHC for $m_{3/2}$ (TeV) in two HCAMSB 
model lines: $\alpha =0.025$ (upper entry) and $\alpha =0.15$ (lower entry),
in various signal channels.
}
\label{tab:reach}
\end{center}
\end{table}
%


\section{Discussion and conclusions}
\label{sec:conclude}

In this paper, we have examined some phenomenological consequences of
hypercharged anomaly-mediated SUSY breaking models at the LHC. We have 
computed the expected sparticle mass spectrum, and mapped out the 
relevant parameter space of the HCAMSB model. We have computed sparticle
branching fractions, production cross sections and expected LHC
collider events, and compared against expectations for SM backgrounds. 
Our main result was to compute the reach of the LHC for
HCAMSB models assuming 100 fb$^{-1}$ of integrated luminosity. We find
an LHC reach to $m_{3/2}\sim 115$ TeV (corresponding to $m_{\tg}\sim 2.4$ TeV)
 for low values of $\alpha$, and a reach to $m_{3/2}\sim 105$ TeV 
(corresponding to $m_{\tg}\sim 2.2$ TeV) for large $\alpha$. 
We expect the reach for $\mu <0$ to be similar to the reach for $\mu >0$, due to
similarities in the spectra for the two cases (see Fig. \ref{fig:m10}.)
We also expect the reach for large $\tan\beta$ to be similar to the reach for
low $\tan\beta$ in the $0\ell$ and $1\ell$ channels (differences in the multi-lepton channels can
occur due to enhanced -ino decays to taus and $b$s at large $\tan\beta$). 
The LHC reach for HCAMSB models tends to be somewhat lower than the reach
for mAMSB models, where Ref. \cite{bmt,barr} finds a 100 fb$^{-1}$ reach of
$m_{\tg}\sim 2.75$  TeV for low values of $m_0$. 
This is due in part because, in mAMSB, the various squark states are more
clustered about a common mass scale $m_0$, while in HCAMSB the squark states
are highly split, with $m_{\tq_R}\gg m_{\tq_L}\sim m_{\tg}$.

The HCAMSB LHC event characteristics suffer similarities and differences with generic
mAMSB models. Both HCAMSB and mAMSB give rise to multi-jet plus 
multi-lepton plus $\eslt$ event topologies, and within these event classes,
it is expected that occasional HITs of length a few {\it cm}
will be found, arising from production of the long-lived wino-like 
chargino states. Some of the major differences between the models
include the following.
\bi
\item A severe left-right splitting of scalar masses
is expected in HCAMSB, while left-right scalar degeneracy tends to occur
in mAMSB. This may be testable if some of the slepton states are 
accessible to LHC searches. It is well known that in mAMSB, 
$m_{\te_L}\simeq m_{\te_R}$, while in HCAMSB, $m_{\te_L}\ll m_{\te_R}$,
since the $\te_R$ state has a large weak hypercharge quantum number.
In addition, the lightest stau state, $\ttau_1$, is expected to be 
mainly a left- state in HCAMSB, while it is mixed, but mainly a right- 
state in mAMSB. While it is conceivable that the left-right mixing
might be determined at LHC (using branching fractions or tau energy
distributions), such measurements would be easily performed at a linear
$e^+e^-$ collider, especially using polarized beams\cite{ray}.
\item In HCAMSB models, the light third generation 
squarks $\tst_1$ and $\tb_1$ are expected to be generically lighter
than the gluino mass, and frequently much lighter. This leads to cascade decays
which produce large multiplicities of $b$ and $t$ quarks in the final state.
Thus, in HCAMSB models, a rather high multiplicity of $b$ jets is expected.
In mAMSB, a much lower mutiplicity of $b$-jets is expected, although
this depends also on the value of $\tan\beta$ which is chosen.
\item In HCAMSB models, the $U(1)$ gaugino mass $M_1$ is expected to be
the largest of the gaugino masses, with a mass hierarchy of 
$M_1>\mu >M_2$. This usually implies that the 
$\tz_4$ neutralino is mainly bino-like, while $\tz_2$ and $\tz_3$ are 
higgsino-like, and $\tz_1$ is wino-like. In contrast, in the mAMSB model, 
usually the ordering is that $\mu >M_1>M_2$, so that while $\tz_1$ is
again wino-like, the $\tz_2$ state is bino-like, and $\tz_3$ and $\tz_4$
are higgsino-like. The compositions of the $\tz_i$ for $i>1$ will
not be easy to determine at LHC, but will be more easily determined
at a linear $e^+e^-$ collider. However, the mass ordering gives rise
to OS dilepton distributions with a prominent $Z\to\ell^+\ell^-$ peak 
in HCAMSB, while such a peak should be largely absent in mAMSB models 
(except at large $\tan\beta$ where there is greater mixing in the neutralino sector).
Thus, cascade decay events containing HITs along with a $Z\to\ell^+\ell^-$
peak in the OS dilepton invariant mass distribution may be a smoking gun 
signature for HCAMSB models at the LHC, at least within the lower range
of $\tan\beta$.
\ei

\section*{Acknowledgments}
This work was supported in part by the U.S.~Department of Energy.
%

%
\end{document}